\documentclass[aps,prd,nofootinbib,amsmath,amssymb,superscriptaddress,twocolumn,10pt]{revtex4}

\pdfoutput=1

\usepackage{graphicx}
\usepackage{dcolumn}
\usepackage{bm}
\usepackage{amssymb}
\usepackage{booktabs}
\usepackage{amsmath}
\usepackage{multirow}
\usepackage[colorlinks=true, linkcolor=purple, citecolor=blue]{hyperref}

\newcommand{\be}{\begin{equation}}
\newcommand{\ee}{\end{equation}}
\newcommand{\bq}{\begin{eqnarray}}
\newcommand{\eq}{\end{eqnarray}}

\usepackage[usenames,dvipsnames]{xcolor}

\begin{document}

\title{Constraints on neutrino mass in the scenario of vacuum energy interacting with cold dark matter after Planck 2018}

\author{Hai-Li Li}
\affiliation{Department of Physics, College of Sciences, Northeastern
University, Shenyang 110819, China}
\author{Jing-Fei Zhang}
\affiliation{Department of Physics, College of Sciences, Northeastern
University, Shenyang 110819, China}
\author{}
\affiliation{Department of Physics, College of Sciences, Northeastern
University, Shenyang 110819, China}

\author{Xin Zhang\footnote{Corresponding author}}
\email{zhangxin@mail.neu.edu.cn}
\affiliation{Department of Physics, College of Sciences, Northeastern
University, Shenyang 110819, China}
\affiliation{Ministry of Education's Key Laboratory of Data Analytics and Optimization
for Smart Industry, Northeastern University, Shenyang 110819, China}
\affiliation{Center for High Energy Physics, Peking University, Beijing 100080, China}
\affiliation{Center for Gravitation and Cosmology, Yangzhou University, Yangzhou 225009, China
}
\date{\today}

\begin{abstract}
In this work, we investigate the constraints on the total neutrino mass in the scenario of vacuum energy interacting with cold dark matter (abbreviated as I$\Lambda$CDM) by using the latest cosmological observations. We consider four typical interaction forms, i.e., $Q=\beta H \rho_{\rm de}$, $Q=\beta H \rho_{\rm c}$, $Q=\beta H_{0} \rho_{\rm de}$, and $Q=\beta H_{0} \rho_{\rm c}$, in the I$\Lambda$CDM scenario. To avoid the large-scale instability problem in interacting dark energy models, we employ the extended parameterized post-Friedmann method for interacting dark energy to calculate the perturbation evolution of dark energy in these models. The observational data used in this work include the cosmic microwave background (CMB) measurements from the Planck 2018 data release, the baryon acoustic oscillation (BAO) data, the type Ia supernovae (SN) observation (Pantheon compilation), and the 2019 local distance ladder measurement of the Hubble constant $H_{0}$ from the Hubble Space Telescope. We find that, compared with those in the $\Lambda$CDM+$\sum m_{\nu}$ model, the constrains on $\sum m_{\nu}$ are looser in the four I$\Lambda$CDM+$\sum m_{\nu}$ models. When considering the three mass hierarchies of neutrinos, the constraints on $\sum m_{\nu}$ are tightest in the degenerate hierarchy case and loosest in the inverted hierarchy case. In addition, in the four I$\Lambda$CDM+$\sum m_{\nu}$ models, the values of coupling parameter $\beta$ are larger using the CMB+BAO+SN+$H_{0}$ data combination than that using the CMB+BAO+SN data combination, and $\beta>0$ is favored at more than 1$\sigma$ level when using CMB+BAO+SN+$H_{0}$ data combination. The issue of the $H_{0}$ tension is also discussed in this paper. We find that, compared with the $\Lambda$CDM+$\sum m_{\nu}$ model, the $H_{0}$ tension can be alleviated in the I$\Lambda$CDM+$\sum m_{\nu}$ model to some extent.
\end{abstract}
\maketitle

\section{Introduction}\label{sec1}

The phenomenon of neutrino oscillation indicates that neutrinos have nonzero masses and there are mass splttings between different neutrino species \cite{Lesgourgues:2006nd,Agashe:2014kda}. The neutrino oscillation experiments can provide the information about the squared mass differences between the neutrino mass eigenstates. Specifically, the solar and reactor experiments give the result of $\Delta m_{21}^{2} \simeq 7.5 \times 10^{-5}$ $\rm eV^{2}$, and the atmospheric and accelerator beam experiments give the result of $|\Delta m_{31}^{2}| \simeq 2.5 \times 10^{-3}$ $\rm eV^{2}$ \cite{Agashe:2014kda,Xing:2019vks}. Therefore, we can get two possible mass hierarchies of the neutrino mass spectrum, i.e., the normal hierarchy (NH) with $m_{1} < m_{2} \ll m_{3}$ and the inverted hierarchy (IH) with $m_{3} \ll m_{1} < m_{2}$, where $m_{1}$, $m_{2}$, and $m_{3}$ denote the masses of neutrinos for the three mass eigenstates. However, the absolute masses of neutrinos are still unknown.

In principle, laboratory experiments of particle physics can directly measure the absolute masses of neutrinos, but these experiments have always been facing great challenges \cite{Osipowicz:2001sq, KlapdorKleingrothaus:2002ip, KlapdorKleingrothaus:2004wj, Kraus:2004zw, Otten:2008zz, Wolf:2008hf, Huang:2016qmh, Zhang:2017ljh, Betts:2013uya}. Compared with these particle physics experiments, cosmological observations are more prone to be capable of measuring the absolute masses of neutrinos \cite{Valle:2005ai,Hannestad:2010kz,Lesgourgues:2012uu}, since massive neutrinos can leave rich signatures on the cosmic microwave background (CMB) anisotropies and the large-scale structure (LSS) formation at different epochs of the cosmic evolution \cite{Abazajian:2013oma}. Thus, we can extract useful information on neutrinos from these available cosmological observations.

Recently, the issue of cosmological constraints on the total neutrino mass with the consideration of mass hierarchy using the latest observational data has been discussed in Ref. \cite{Zhang:2020mox}. In Ref. \cite{Zhang:2020mox}, the authors discussed the constraints on neutrino mass in several typical dark energy models, e.g., the $\Lambda$ cold dark matter ($\Lambda$CDM), $w$CDM, Chevallier-Polarski-Linder (CPL), and holographic dark energy (HDE) models. It was found that, compared to the $\Lambda$CDM+$\sum m_{\nu}$ model, larger neutrino masses are favored in the $w$CDM+$\sum m_{\nu}$ and CPL+$\sum m_{\nu}$ models, and the most stringent upper limits are obtained in the HDE+$\sum m_{\nu}$ model. Moreover, in Ref. \cite{Zhang:2020mox}, it was also confirmed that the NH case is more favored by current cosmological observations than the IH case. For more relevant studies on constraining the total neutrino mass by using cosmological observations, see e.g., Refs. \cite{Hu:1997mj,Reid:2009nq,Thomas:2009ae,Carbone:2010ik,Li:2012vn,Wang:2012vh,Wang:2012uf,Audren:2012vy,Riemer-Sorensen:2013jsa,Font-Ribera:2013rwa,Zhang:2014dxk,Zhang:2014nta,Zhang:2014ifa,Palanque-Delabrouille:2014jca,Geng:2014yoa,Li:2015poa,Ade:2015xua,Zhang:2015rha,Geng:2015haa,Chen:2015oga,Allison:2015qca,Cuesta:2015iho,Chen:2016eyp,Moresco:2016nqq,Lu:2016hsd,Kumar:2016zpg,Xu:2016ddc,Vagnozzi:2017ovm,Zhang:2017rbg,Lorenz:2017fgo,Zhao:2017jma,Vagnozzi:2018jhn,Wang:2018lun,Li:2017iur,Wang:2017htc,Feng:2017usu,Zhao:2016ecj,Zhang:2015uhk,Huang:2015wrx, Wang:2016tsz,Vagnozzi:2019utt,Vagnozzi:2019jzi,Giusarma:2016phn,Gariazzo:2018pei,Liu:2020vgn,Choudhury:2018byy,Allahverdi:2016fvl,Han:2017wnk,Zhou:2014fva,Huo:2011ve,Zhang:2019ipd,DiazRivero:2019ukx}.

Furthermore, the impacts of interaction between dark energy (DE) and cold dark matter (CDM) on constraining neutrino mass have also been considered. For example, in the scenario of vacuum energy interacting with cold dark matter, which is abbreviated as the I$\Lambda$CDM scenario in this work, the constraint on $\sum m_{\nu}$ becomes $\sum m_{\nu} < 0.10$ eV (2$\sigma$) for $Q=\beta H \rho_{\rm de}$, $\sum m_{\nu} < 0.20$ eV (2$\sigma$) for $Q=\beta H \rho_{\rm c}$ \cite{Guo:2017hea}, and $\sum m_{\nu} < 0.214$ eV (2$\sigma$) for $Q=\beta H_{0} \rho_{\rm c}$ \cite{Feng:2019jqa}. When the mass hierarchies of neutrinos are considered in the I$\Lambda$CDM model \cite{Guo:2018gyo, Feng:2019mym}, the results showed that the degenerate hierarchy (DH) case gives the smallest upper limit of the neutrino mass and the NH case is more favored over the IH case. In the present work, we will revisit the constraints on the total neutrino mass in the I$\Lambda$CDM scenario after the Planck 2018 data release. We will consider more forms of interaction term $Q$, and also adopt the mass hierarchies of neutrinos in this work.

In the so-called ``interacting dark energy" (IDE) scenario, some direct, non-gravitational coupling between dark energy and dark matter is assumed and its cosmological consequences have been widely studied \cite{Amendola:1999er,Amendola:2001rc,Comelli:2003cv, Cai:2004dk, Zhang:2005rg, Zimdahl:2005bk, Zhang:2004gc, Wang:2006qw, Guo:2007zk, Bertolami:2007zm, Zhang:2007uh, Boehmer:2008av, Valiviita:2008iv, He:2008tn, He:2009mz, He:2009pd, Koyama:2009gd, Xia:2009zzb, Li:2009zs, Zhang:2009qa, Wei:2010cs, Li:2010ak, He:2010im, Li:2011ga, Fu:2011ab, Zhang:2012uu, Zhang:2013lea, Li:2013bya, Geng:2015ara,Yin:2015pqa, Murgia:2016ccp,Wang:2016lxa, Pourtsidou:2016ico, Costa:2016tpb, Sola:2016ecz, Feng:2016djj, Xia:2016vnp, vandeBruck:2016hpz, Sola:2016zeg, Kumar:2017dnp, Sola:2017jbl}. Theoretically speaking, the consideration of such an interaction is helpful in solving the cosmic coincidence problem \cite{Comelli:2003cv,Cai:2004dk,Zhang:2005rg,He:2008tn,He:2009pd}, but actually what is more important is to detect such an interaction using the cosmological observations. The impacts of interactions between dark energy and dark matter on the CMB \cite{He:2009pd,Pourtsidou:2016ico} and LSS \cite{Amendola:2001rc,Bertolami:2007zm,He:2008tn,Koyama:2009gd,Li:2013bya,Pourtsidou:2016ico} have been studied in-depth.


In this paper, we only consider the simplest class of models in the IDE scenario, i.e., the I$\Lambda$CDM models, in which the vacuum energy with $w=-1$ serves as dark energy. In this scenario, the energy conservation equations of the vacuum energy and the cold dark matter satisfy
\begin{equation}\label{conservation1}
\dot{\rho_{\rm de}} = Q,
\end{equation}
\begin{equation}\label{conservation2}
\dot{\rho_{\rm c}} = -3 H \rho_{\rm c}-Q,
\end{equation}
where $\rho_{\rm de}$ and $\rho_{\rm c}$ represent the densities of dark energy (namely, vacuum energy) and cold dark matter, respectively, $H$ is the Hubble parameter, the dot represents the derivative with respect to the cosmic time $t$, and $Q$ is the energy transfer rate. Usually, the form of $Q$ is assumed to be proportional to the density of dark energy or dark matter, i.e., $Q=\beta H \rho_{\rm de}$ or $Q=\beta H \rho_{\rm c}$, where the appearance of $H$ is only for mathematical convenience. In the research area of interacting dark energy, another perspective is to consider $Q=\beta H_{0} \rho_{\rm de}$ or $Q=\beta H_{0} \rho_{\rm c}$ \cite{Valiviita:2008iv}, where the appearance of $H_{0}$ is only for a dimensional consideration. From Eqs.~(\ref{conservation1}) and ~(\ref{conservation2}), it is known that $\beta > 0$ means cold dark matter decaying into dark energy, $\beta < 0$ means dark energy decaying into cold dark matter, and $\beta = 0$ indicates no interaction between vacuum energy and cold dark matter.

Different phenomenological models of I$\Lambda$CDM can be built by assuming different forms of $Q$. In this work, we will collect the popular forms of $Q$ in the current literature and then focus on the impacts of different forms of $Q$ on constraining the total neutrino mass after the Planck 2018 data release. We will consider the four typical forms of $Q$: $Q=\beta H \rho_{\rm de}$, $Q=\beta H \rho_{\rm c}$, $Q=\beta H_{0} \rho_{\rm de}$ and $Q=\beta H_{0} \rho_{\rm c}$. The mass hierarchies of neutrinos are also considered in this work. In addition, we also wish to see whether some hint of the existence of nonzero interaction can be found in these I$\Lambda$CDM models by using the latest observational data.

This paper is organized as follows. In Sec.~\ref{sec2}, we introduce the cosmological observations used in this work and briefly describe the analysis method. In Sec.~\ref{sec3}, we report the constraint results and then make some relevant discussions. The issue of $H_{0}$ tension will also be discussed in this section. Conclusion is given in Sec.~\ref{sec4}.

\section{Method and data}\label{sec2}


In the I$\Lambda$CDM model, there are seven basic cosmological parameters  $\{\omega_b,~\omega_c,~100\theta_{\rm MC},~\tau,~n_s,~\ln (10^{10}A_s),\beta\}$, where $\omega_b$ is the present density of baryons, $\omega_c$ is the present density of cold dark matter, $\theta_{\rm MC}$ is the ratio between the sound horizon to the angular diameter distance at the decoupling epoch, $\tau$ is the Thomson scattering optical depth to reionization, $n_s$ is the scalar spectral index, $A_s$ is the amplitude of primordial scalar perturbation power spectrum, and $\beta$ is the dimensionless coupling constant describing the coupling strength between vacuum energy and dark matter.

For the I$\Lambda$CDM model there is a problem of early-time perturbation instability, because in the IDE models, the cosmological perturbations of dark energy will be divergent in a part of the parameter space, which ruins the IDE cosmology in the perturbation level. The origin of the difficulty is that we know little about the nature of dark energy, so we do not know how to treat the spread of sounds in dark energy fluid which has a negative equation of state (EoS). To overcome the problem of perturbation instability, in 2014, Yun-He Li, Jing-Fei Zhang, and Xin Zhang established an effective theoretical framework for IDE cosmology based on the extended version of the parameterized post-Friedmann (PPF) approach, which can safely calculate the cosmological perturbations in the whole parameter space of an interacting dark energy model. About the extended PPF method, see Refs. \cite{Li:2014eha,Li:2014cee,Li:2015vla,Zhang:2017ize,Feng:2018yew}, and the original PPF method is introduced in Refs. \cite{Hu:2008zd,Fang:2008sn}. In this work, we will employ the extended PPF method \cite{Li:2014eha,Li:2014cee,Li:2015vla,Zhang:2017ize,Feng:2018yew} to calculate the cosmological perturbations in the I$\Lambda$CDM model.

We use the modified version of the publicly available Markov-Chain Monte Carlo package CosmoMC \cite{Lewis:2002ah} to constrain the neutrino mass and other cosmological parameters. We monitor the convergence of the generated MCMC chains by using the Gelman-Rubin parameter $R$ \cite{Gelman:1992zz}, requiring $R-1< 0.01$ for our MCMC chains to be considered as converged. When considering the neutrino mass splitting, we should note the following rules. For the NH case, the neutrino mass spectrum is
\begin{equation}\label{neuNH}
(m_{1},m_{2},m_{3})=(m_{1},\sqrt{m_{1}^{2}+\Delta m_{21}^{2}},\sqrt{m_{1}^{2}+|\Delta m_{31}^{2}|})
\end{equation}
where $m_{1}$ is a free parameter; for the IH case, the neutrino mass spectrum is
\begin{small}
\begin{equation}\label{neuIH}
(m_{1},m_{2},m_{3})=(\sqrt{m_{3}^{2}+|\Delta m_{31}^{2}|},\sqrt{m_{3}^{2}+|\Delta m_{31}^{2}|+\Delta m_{21}^{2}},m_{3})
\end{equation}
\end{small}
where $m_{3}$ is a free parameter; for comparison, the DH case is also considered, in which the neutrino mass spectrum is
\begin{equation}\label{neuDH}
m_{1}=m_{2}=m_{3}=m
\end{equation}
where $m$ is a free parameter. Note also that the input lower bounds of $\sum m_{\nu}$ are 0.06 eV for the NH case, 0.10 ev for the IH case, and 0 eV for the DH case, respectively.

The current observational data sets we used in this paper include CMB, BAO, SN and $H_{0}$. For the CMB data, we use the Planck TT, TE, EE spectra at $\ell \geq 30$, the low-$\ell$ temperature Commander likelihood, and the low-$\ell$ SimAll EE likelihood, from the Planck 2018 data release \cite{Aghanim:2018eyx}. For the BAO data, we consider the measurements from 6dFGS ($z_{\rm eff}=0.106$) \cite{Beutler:2011hx}, SDSS-MGS ($z_{\rm eff}=0.15$) \cite{Ross:2014qpa}, and BOSS DR12 ($z_{\rm eff}=0.38$, 0.51, and 0.61) \cite{Alam:2016hwk}. For the SN data, we employ the latest Pantheon sample, which is comprised of 1048 data points from the Pantheon compilation \cite{Scolnic:2017caz}. For the $H_{0}$ data, we use the 2019 local distance ladder measurement of the Hubble constant $H_{0}=74.03\pm1.42~ \rm{km~ s^{-1} Mpc^{-1}}$ from the Hubble Space Telescope \cite{Riess:2019cxk}. In our analysis, we will use two data combinations, i.e., CMB+BAO+SN and CMB+BAO+SN+$H_{0}$, to constrain the cosmological parameters.

\begin{figure*}[!htp]
\includegraphics[scale=0.55]{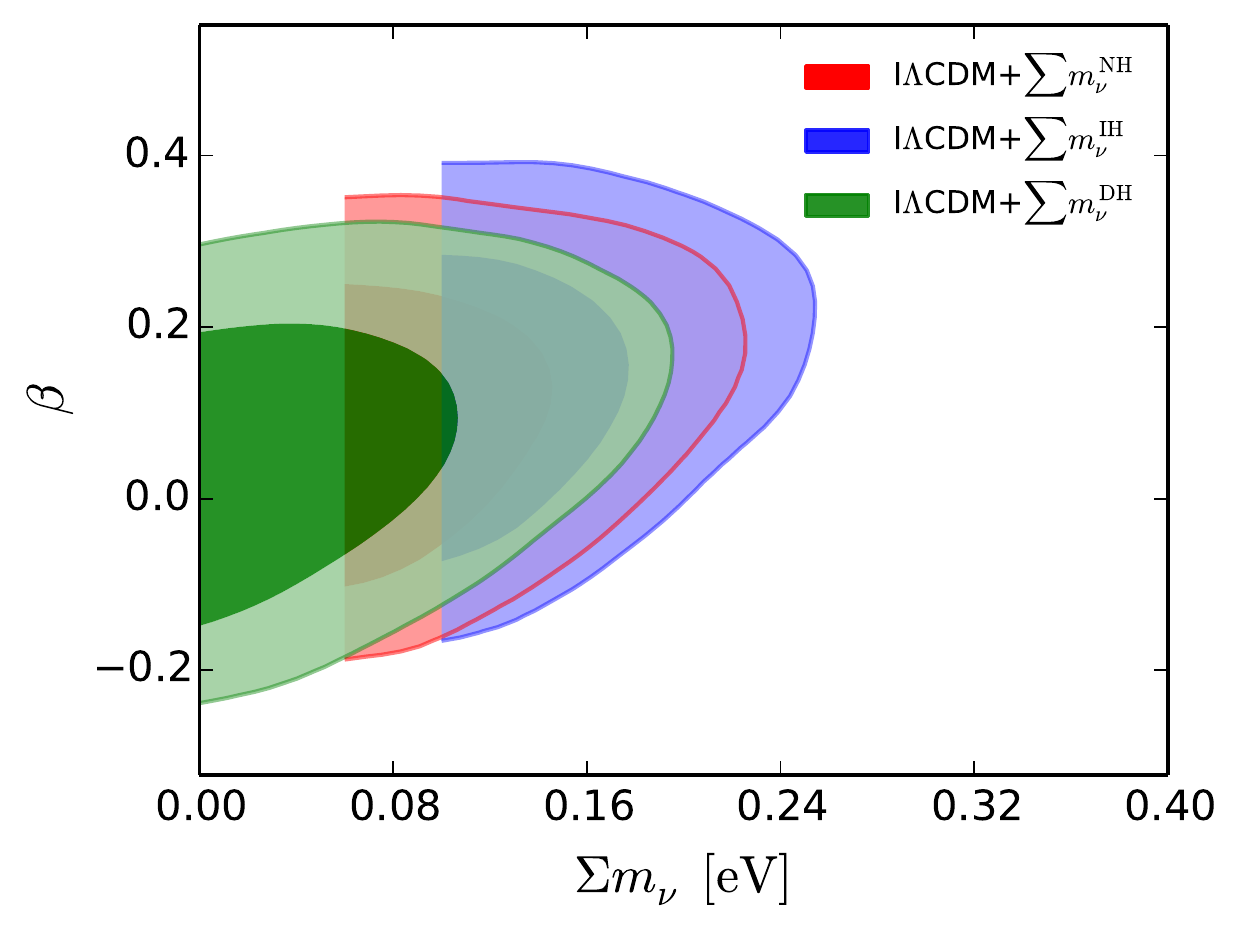}\ \hspace{1cm}
\includegraphics[scale=0.55]{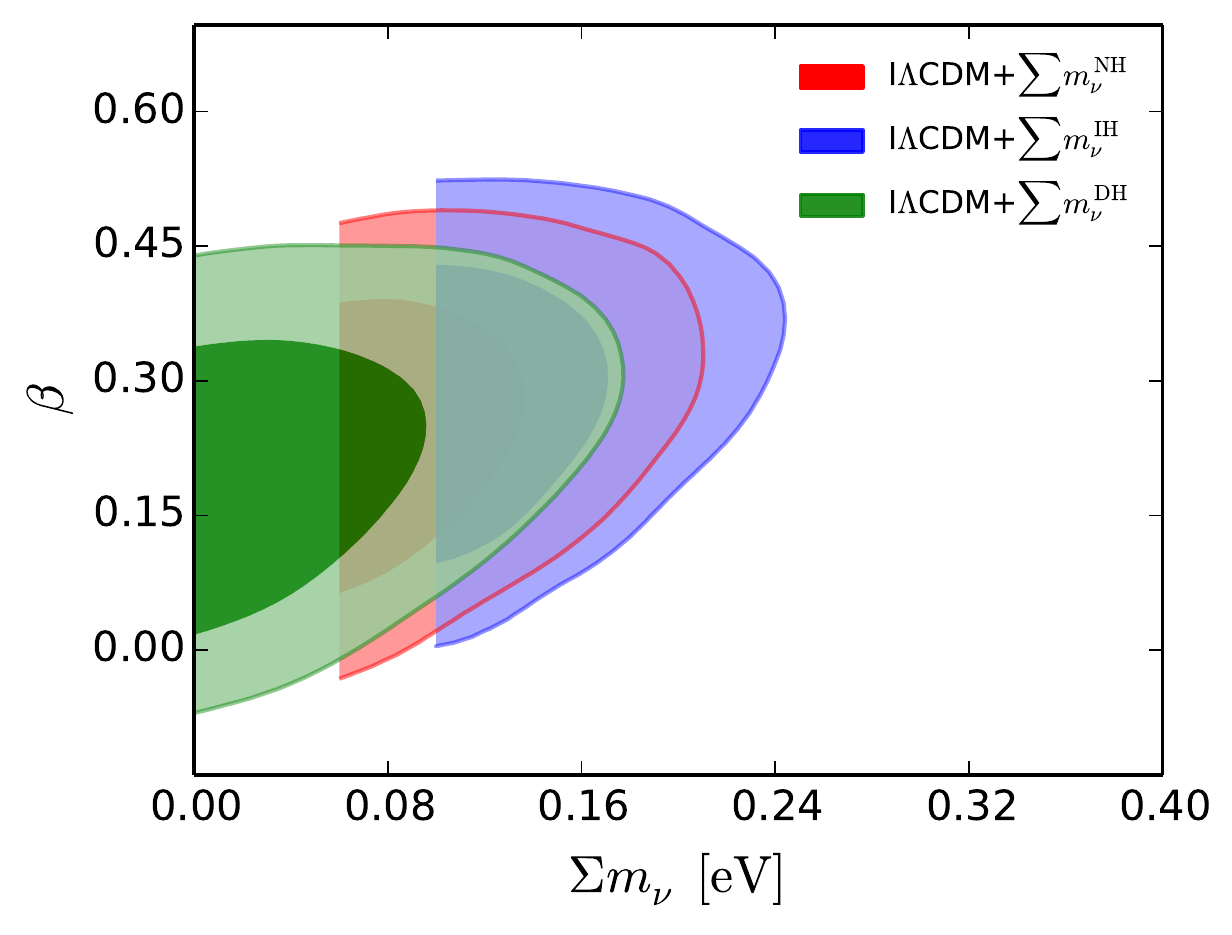}\ \hspace{1cm}
\centering \caption{\label{fig1} Observational constraints (68.3\% and 95.4\% confidence level) on the I$\Lambda$CDM1+$\sum m_{\nu}$ ($Q=\beta H \rho_{\rm de}$) model by using the CMB+BAO+SN (left) and CMB+BAO+SN+$H_{0}$ (right) data combinations, respectively.}
\end{figure*}
\begin{figure*}[!htp]
\includegraphics[scale=0.55]{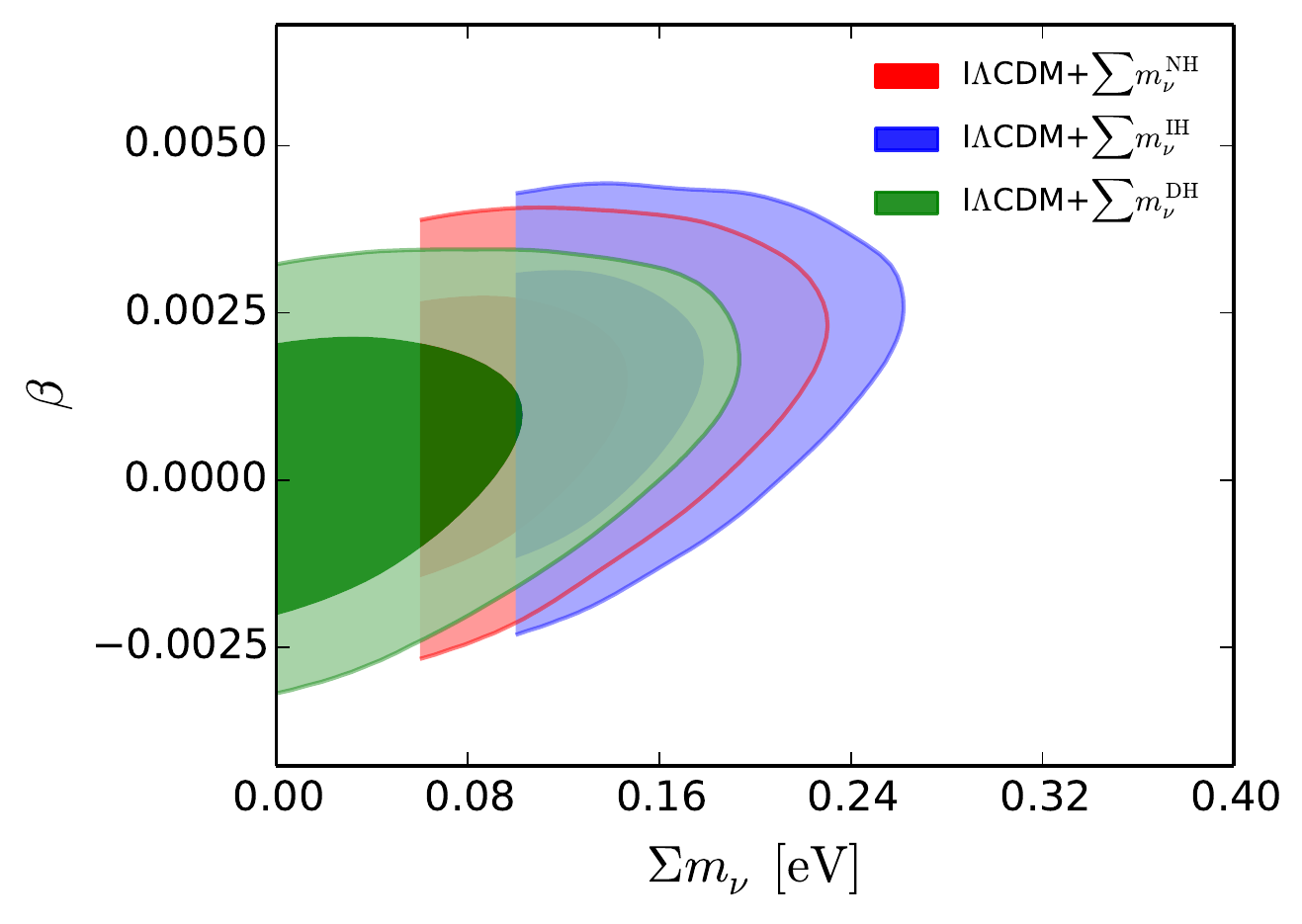}\ \hspace{1cm}
\includegraphics[scale=0.55]{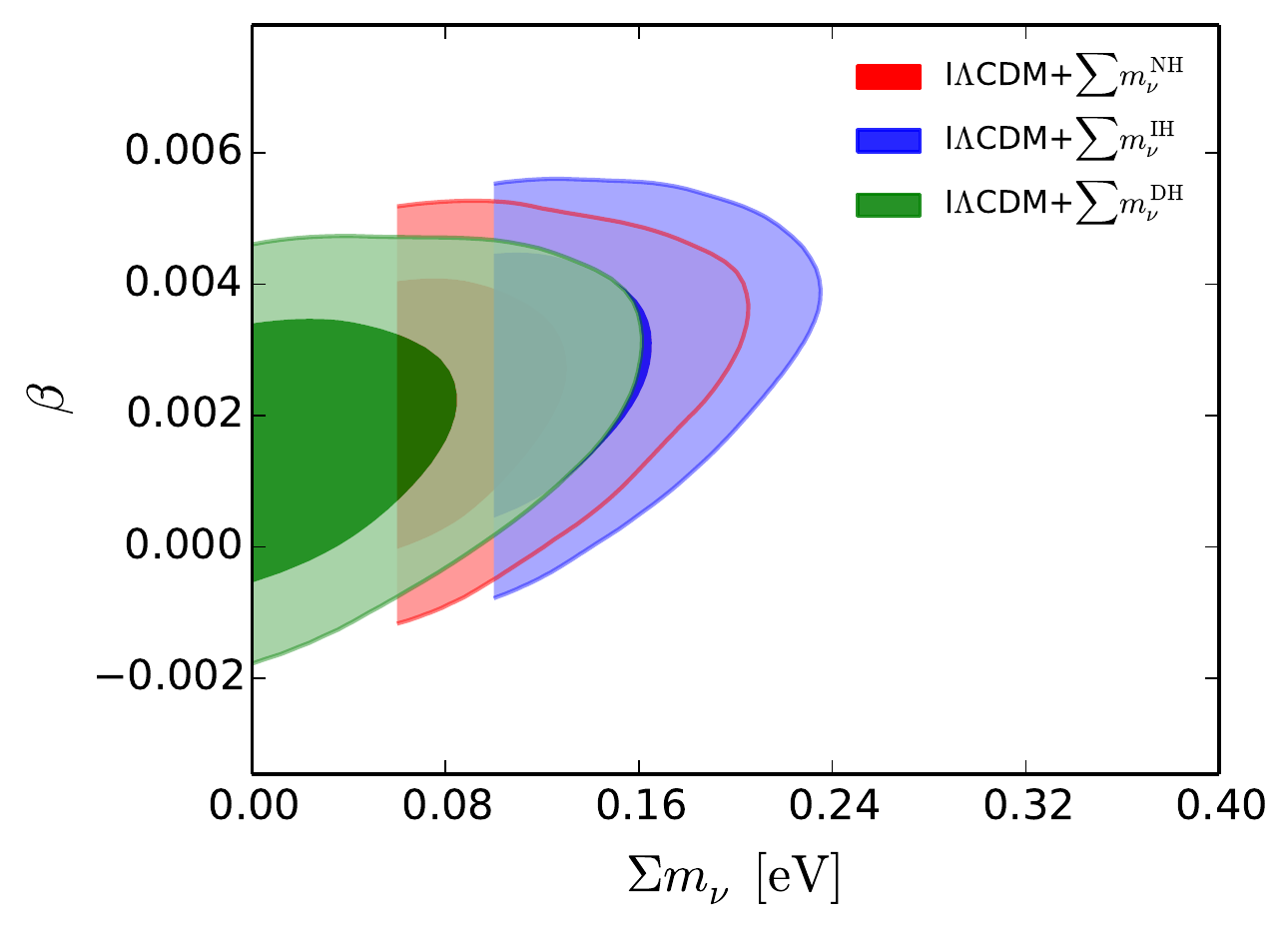}\ \hspace{1cm}
\centering \caption{\label{fig2} Observational constraints (68.3\% and 95.4\% confidence level) on the I$\Lambda$CDM2+$\sum m_{\nu}$ ($Q=\beta H \rho_{\rm c}$) model by using the CMB+BAO+SN (left) and CMB+BAO+SN+$H_{0}$ (right) data combinations, respectively.}
\end{figure*}
\begin{figure*}[!htp]
\includegraphics[scale=0.55]{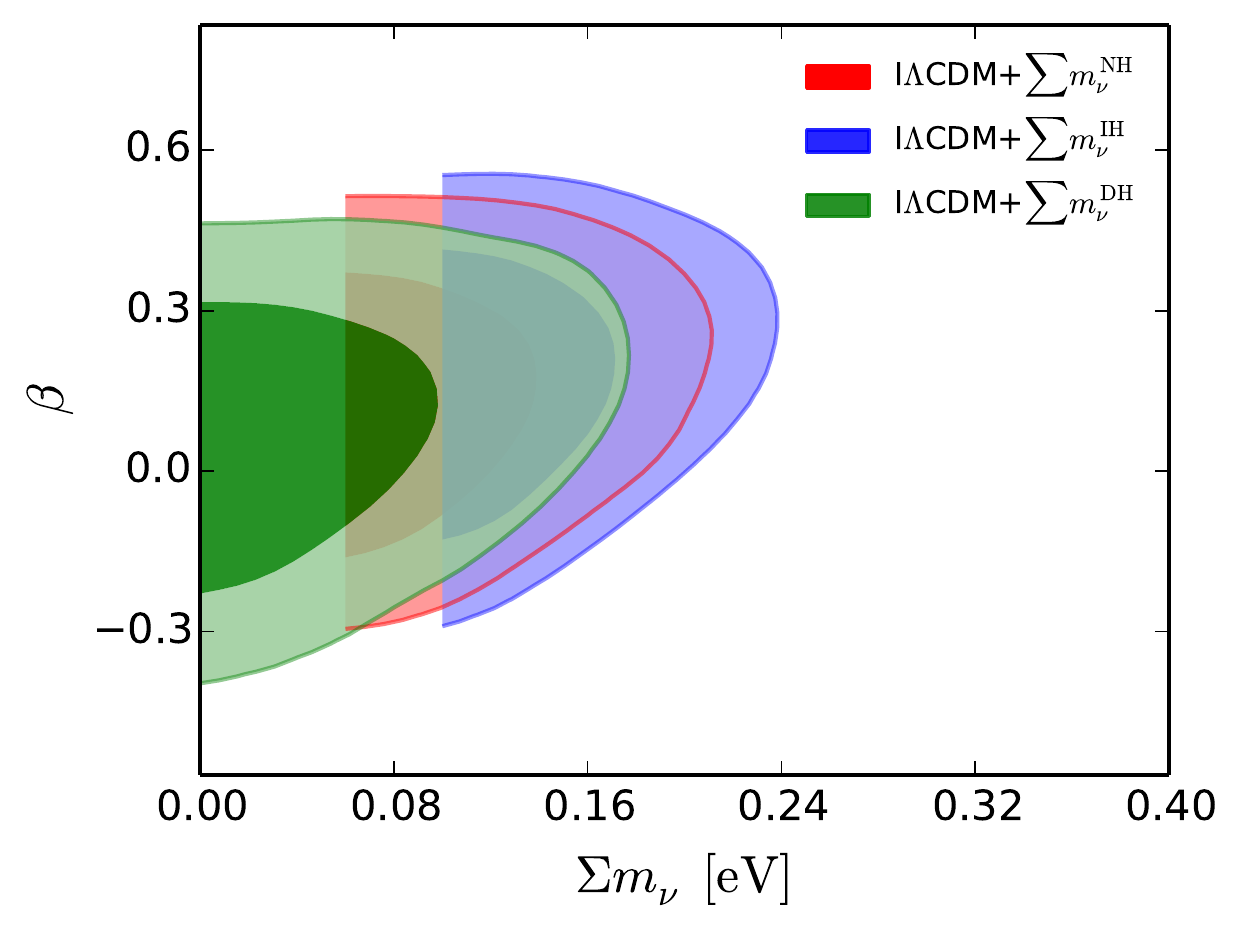}\ \hspace{1cm}
\includegraphics[scale=0.55]{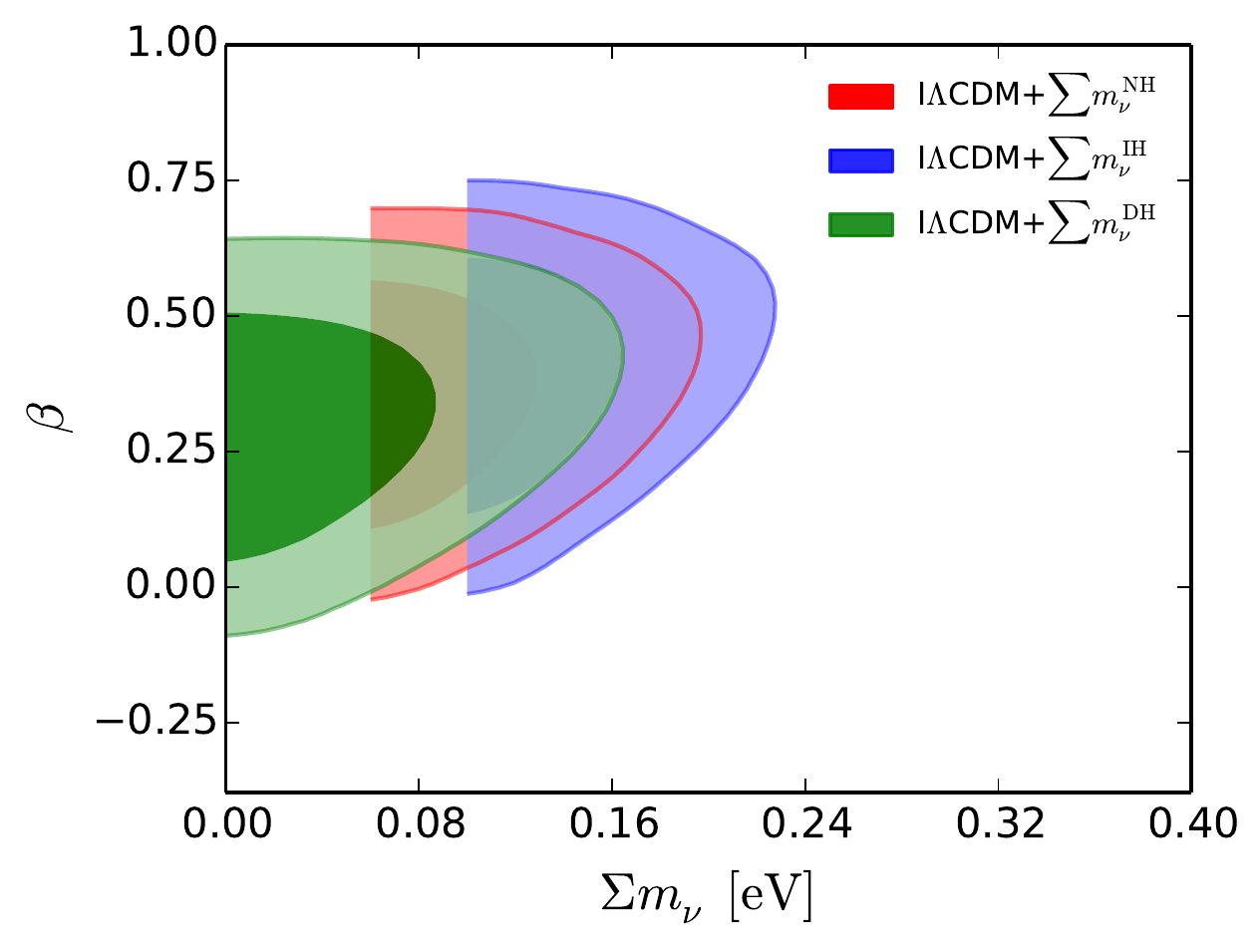}\ \hspace{1cm}
\centering \caption{\label{fig3} Observational constraints (68.3\% and 95.4\% confidence level) on the I$\Lambda$CDM3+$\sum m_{\nu}$ ($Q=\beta H_{0} \rho_{\rm de}$) model by using the CMB+BAO+SN (left) and CMB+BAO+SN+$H_{0}$ (right) data combinations, respectively.}
\end{figure*}
\begin{figure*}[!htp]
\includegraphics[scale=0.55]{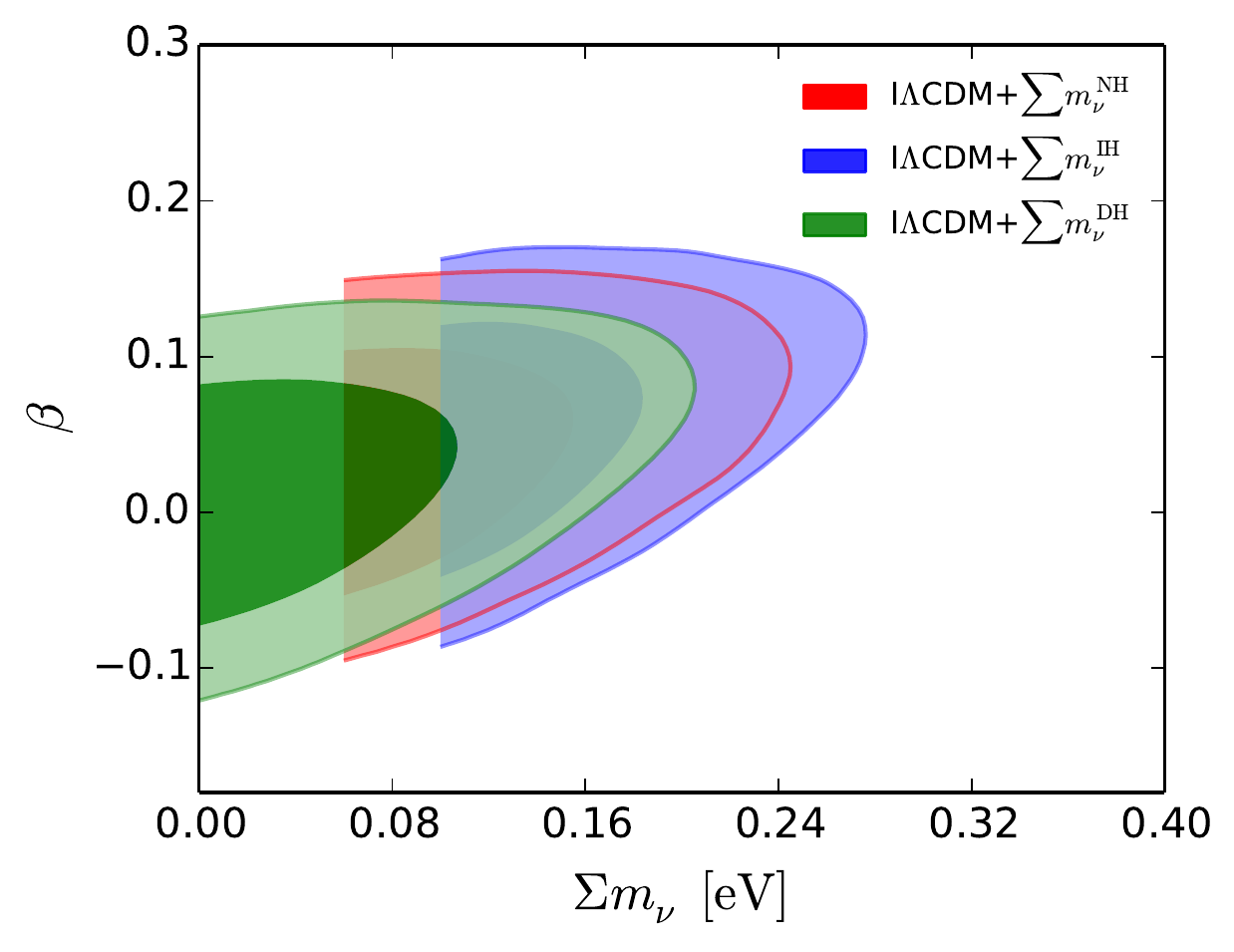}\ \hspace{1cm}
\includegraphics[scale=0.55]{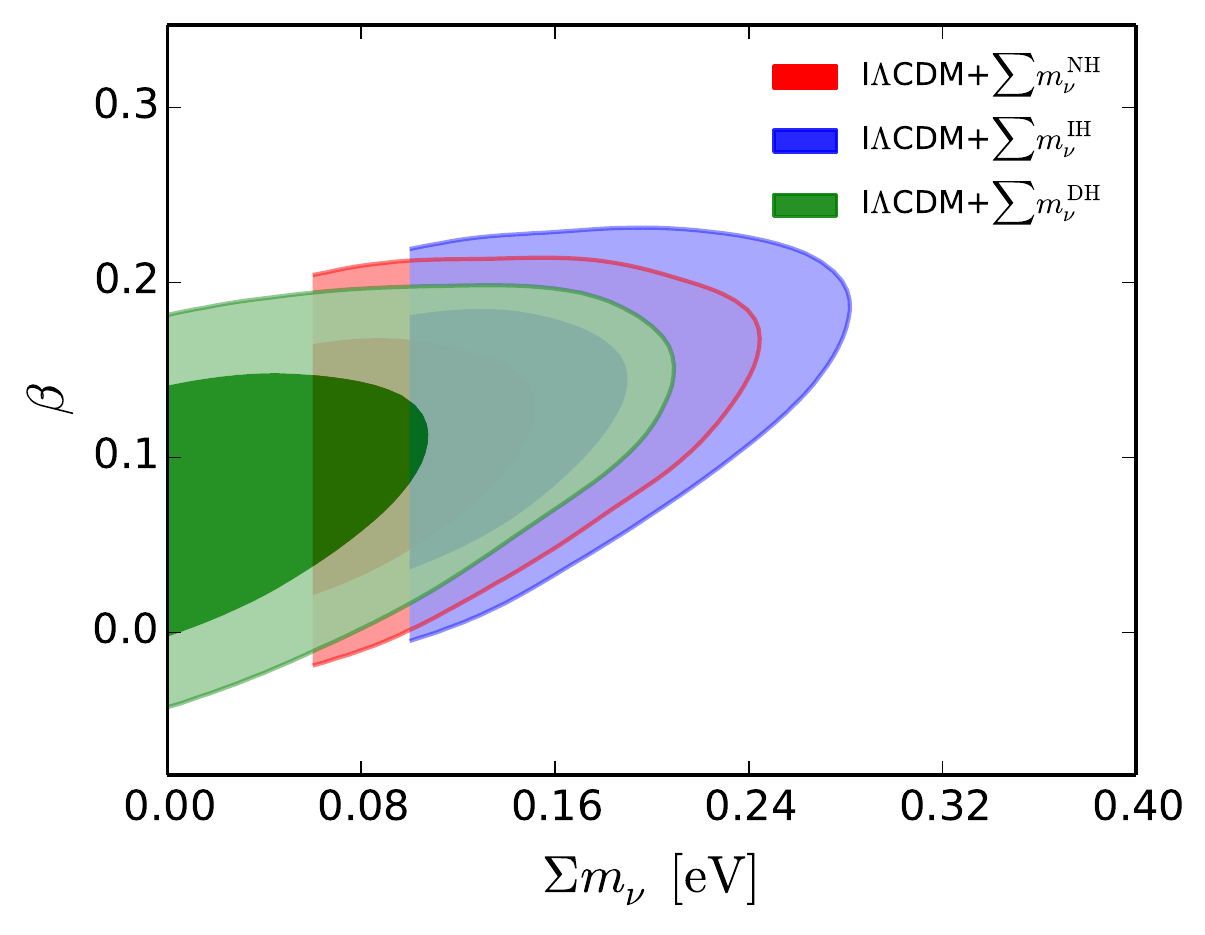}\ \hspace{1cm}
\centering \caption{\label{fig4} Observational constraints (68.3\% and 95.4\% confidence level) on the I$\Lambda$CDM4+$\sum m_{\nu}$ ($Q=\beta H_{0} \rho_{\rm c}$) model by using the CMB+BAO+SN (left) and CMB+BAO+SN+$H_{0}$ (right) data combinations, respectively.}
\end{figure*}

\begin{figure*}[!htp]
\includegraphics[scale=0.3]{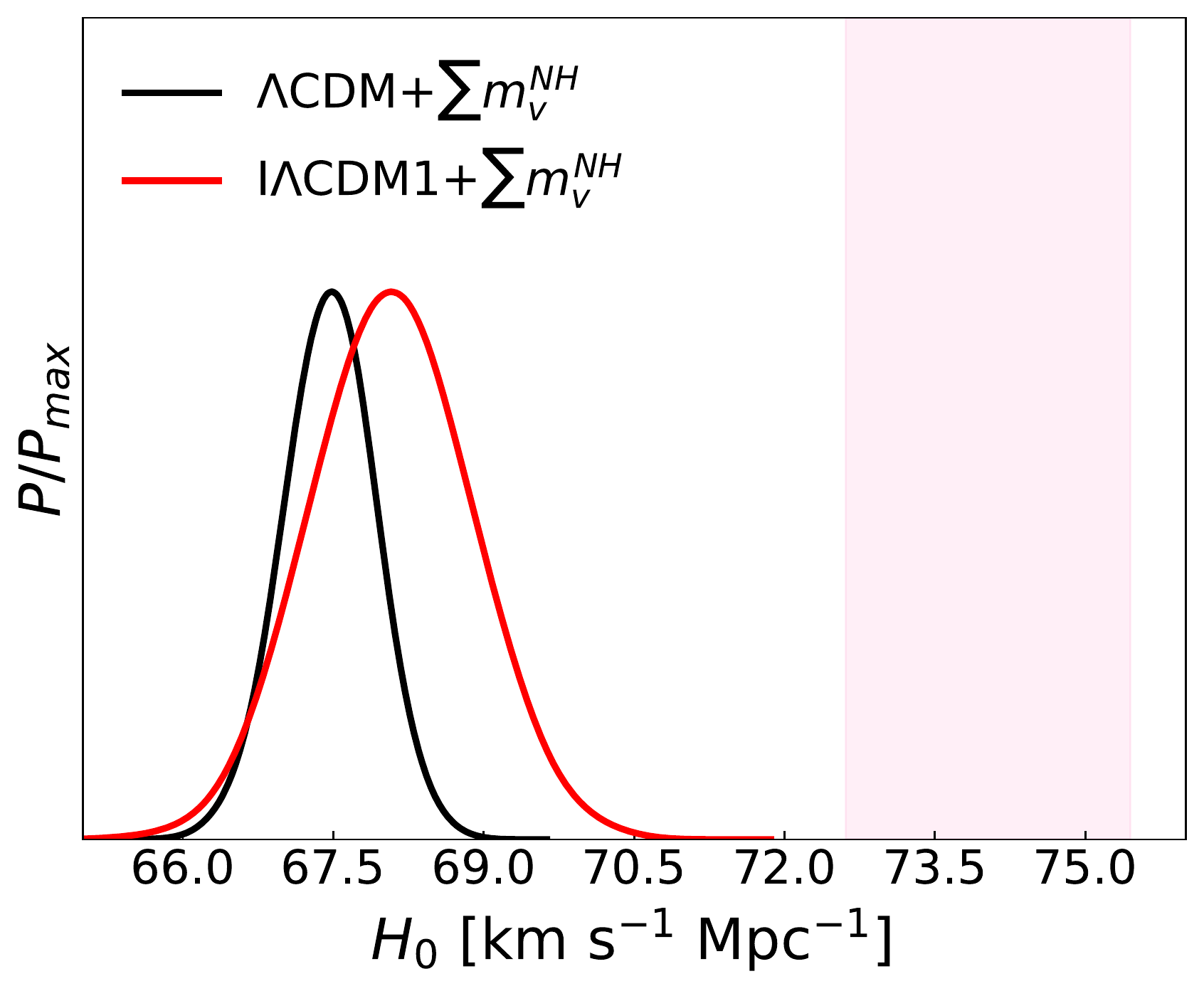}\ \hspace{0.6cm}
\includegraphics[scale=0.3]{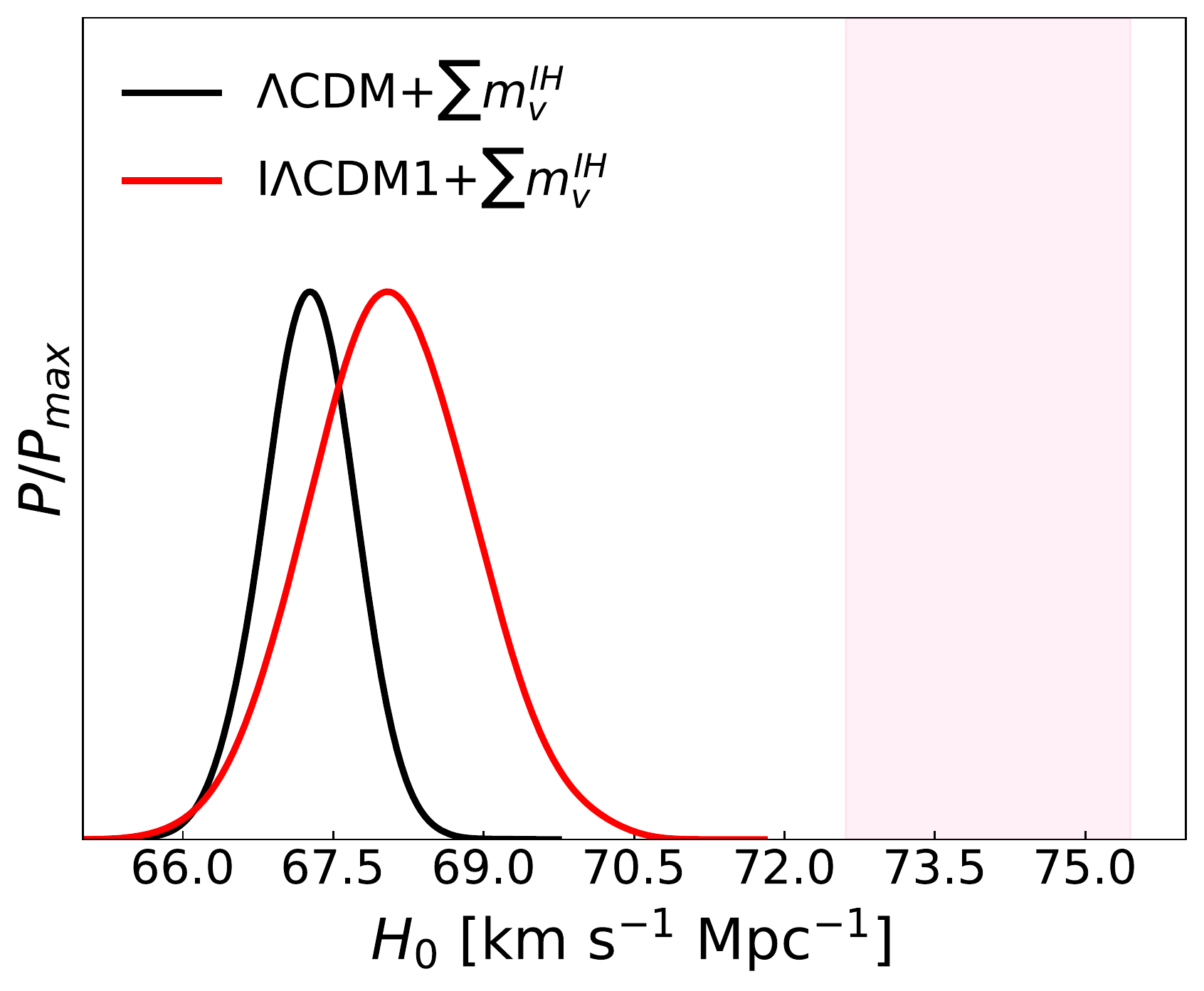}\ \hspace{0.6cm}
\includegraphics[scale=0.3]{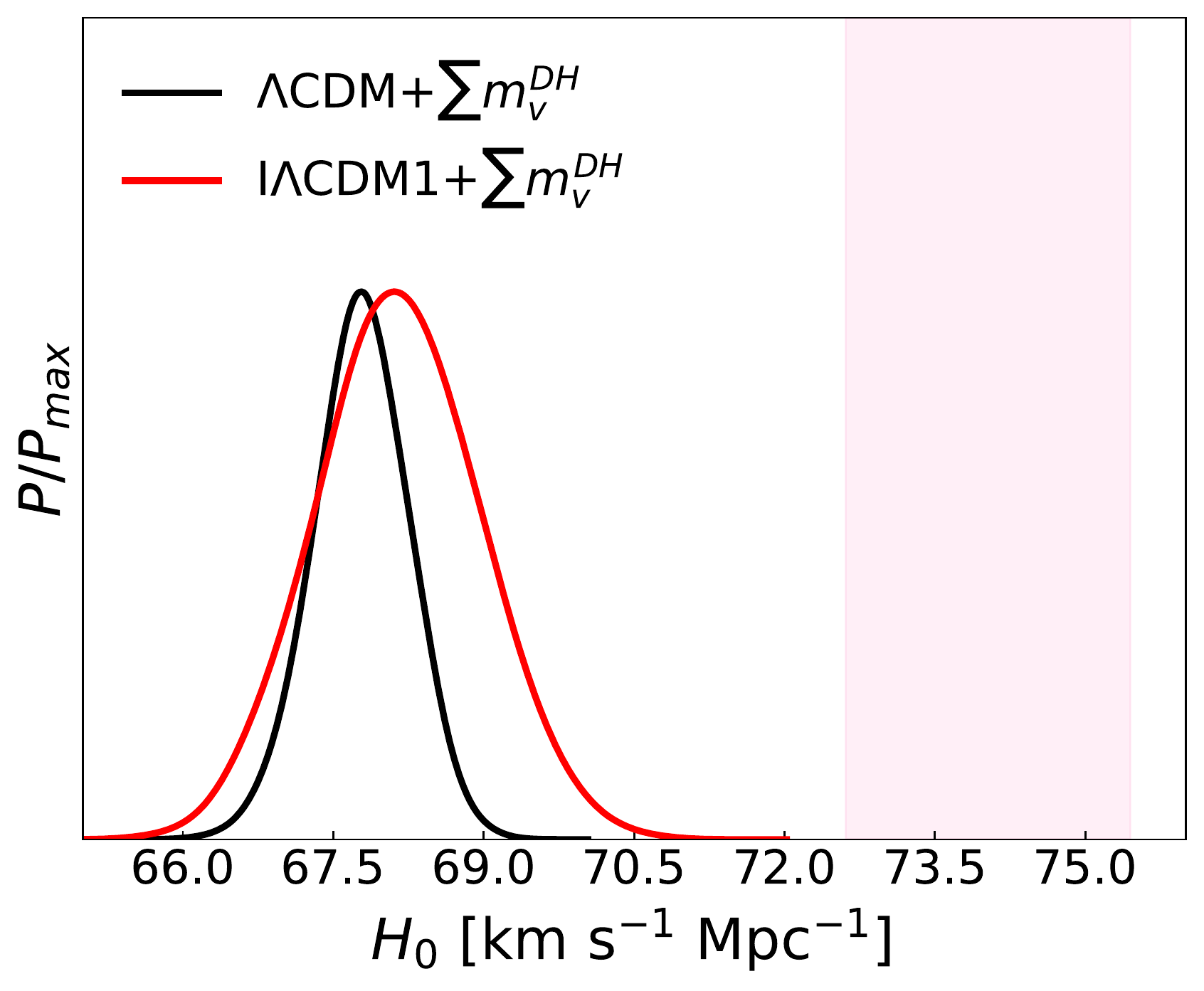}\ \hspace{0.6cm}
\centering \caption{\label{fig5} The one-dimensional posterior distributions for the parameter $H_{0}$ for the $\Lambda$CDM+$\sum m_{\nu}$ (NH, IH, and DH) model and I$\Lambda$CDM1+$\sum m_{\nu}$ (NH, IH, and DH) model with $Q=\beta H \rho_{\rm de}$ by using the CMB+BAO+SN data combination. The result of the local measurement of Hubble constant ($H_{0}=74.03\pm1.42~ \rm{km~ s^{-1} Mpc^{-1}}$) is shown by the hotpink band.}
\end{figure*}

\begin{figure*}[!htp]
\includegraphics[scale=0.3]{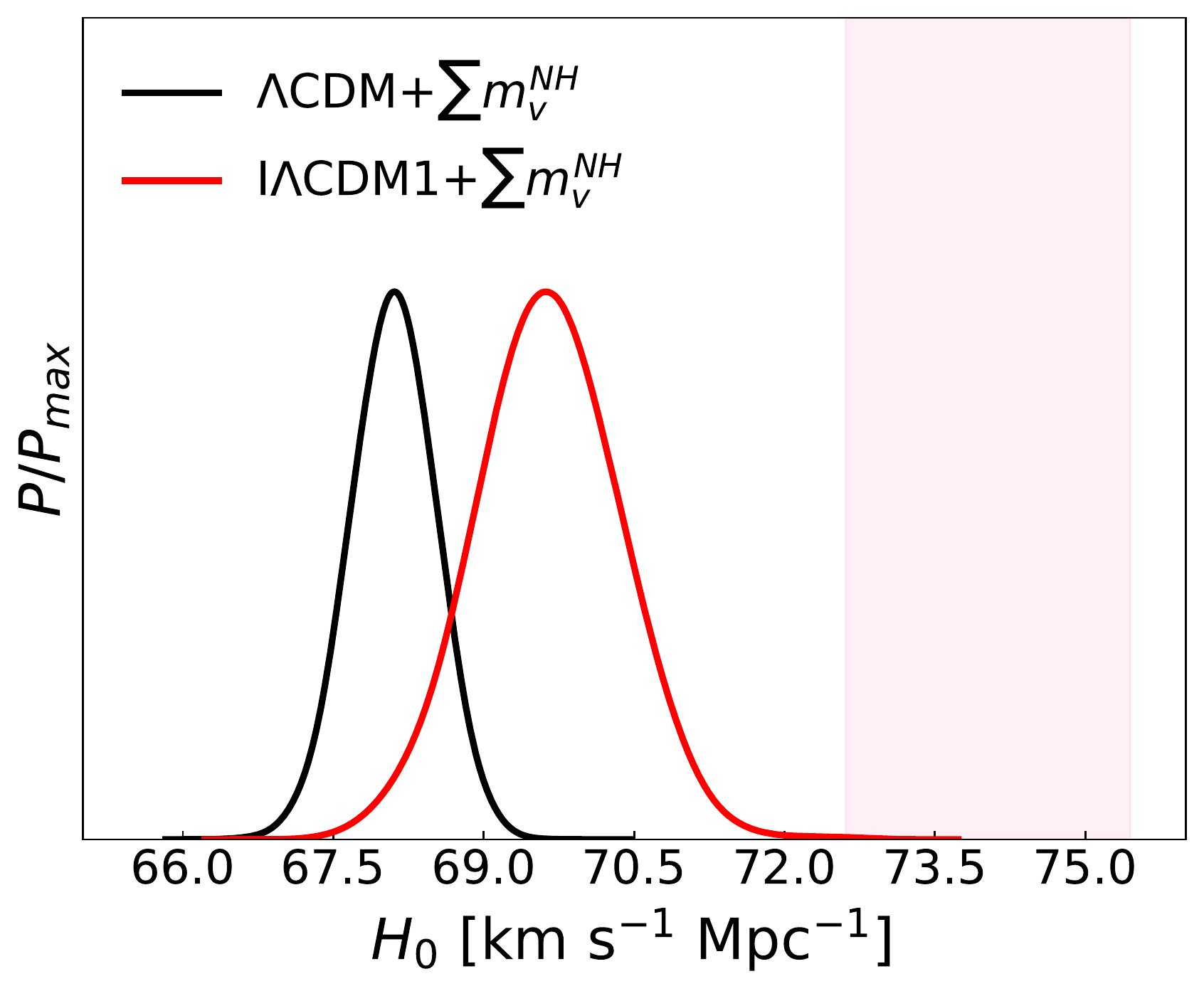}\ \hspace{0.6cm}
\includegraphics[scale=0.3]{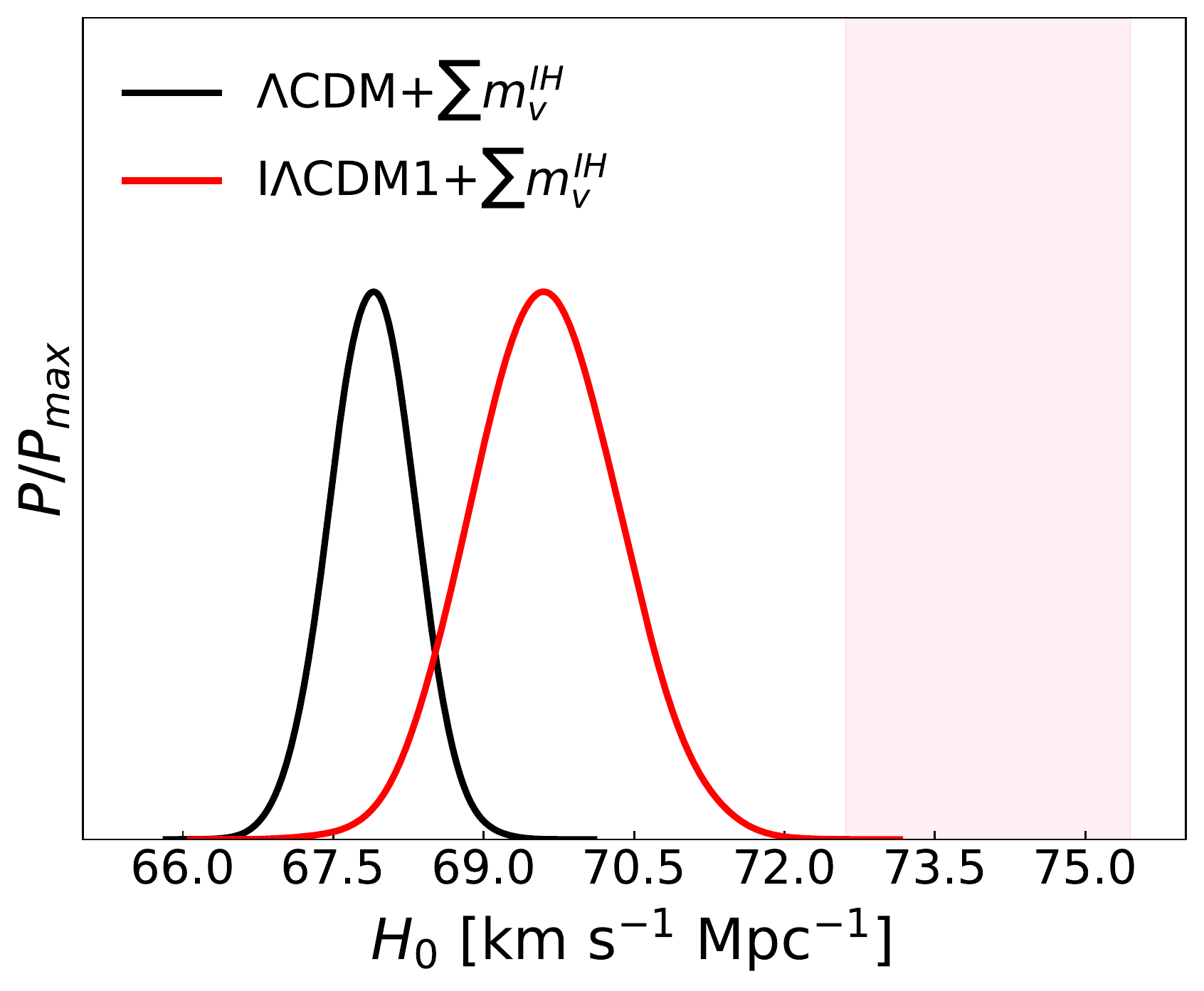}\ \hspace{0.6cm}
\includegraphics[scale=0.3]{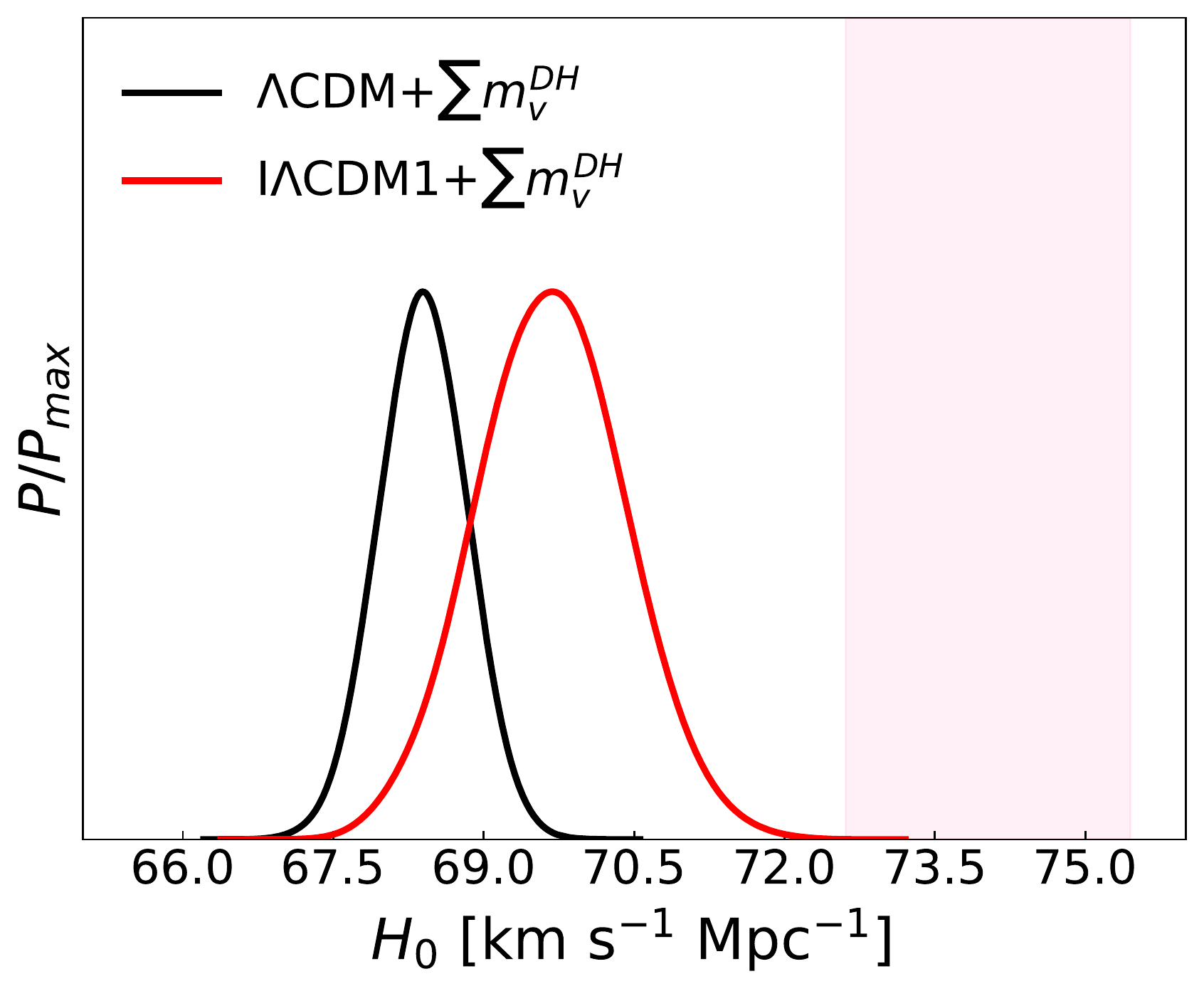}\ \hspace{0.6cm}
\centering \caption{\label{fig6} The one-dimensional posterior distributions for the parameter $H_{0}$ for the $\Lambda$CDM+$\sum m_{\nu}$ (NH, IH, and DH) model and I$\Lambda$CDM1+$\sum m_{\nu}$ (NH, IH, and DH) model with $Q=\beta H \rho_{\rm de}$ by using the CMB+BAO+SN+$H_{0}$ data combination. The result of the local measurement of Hubble constant ($H_{0}=74.03\pm1.42~ \rm{km~ s^{-1} Mpc^{-1}}$) is shown by the hotpink band.}
\end{figure*}

\begin{figure*}[!htp]
\includegraphics[scale=0.3]{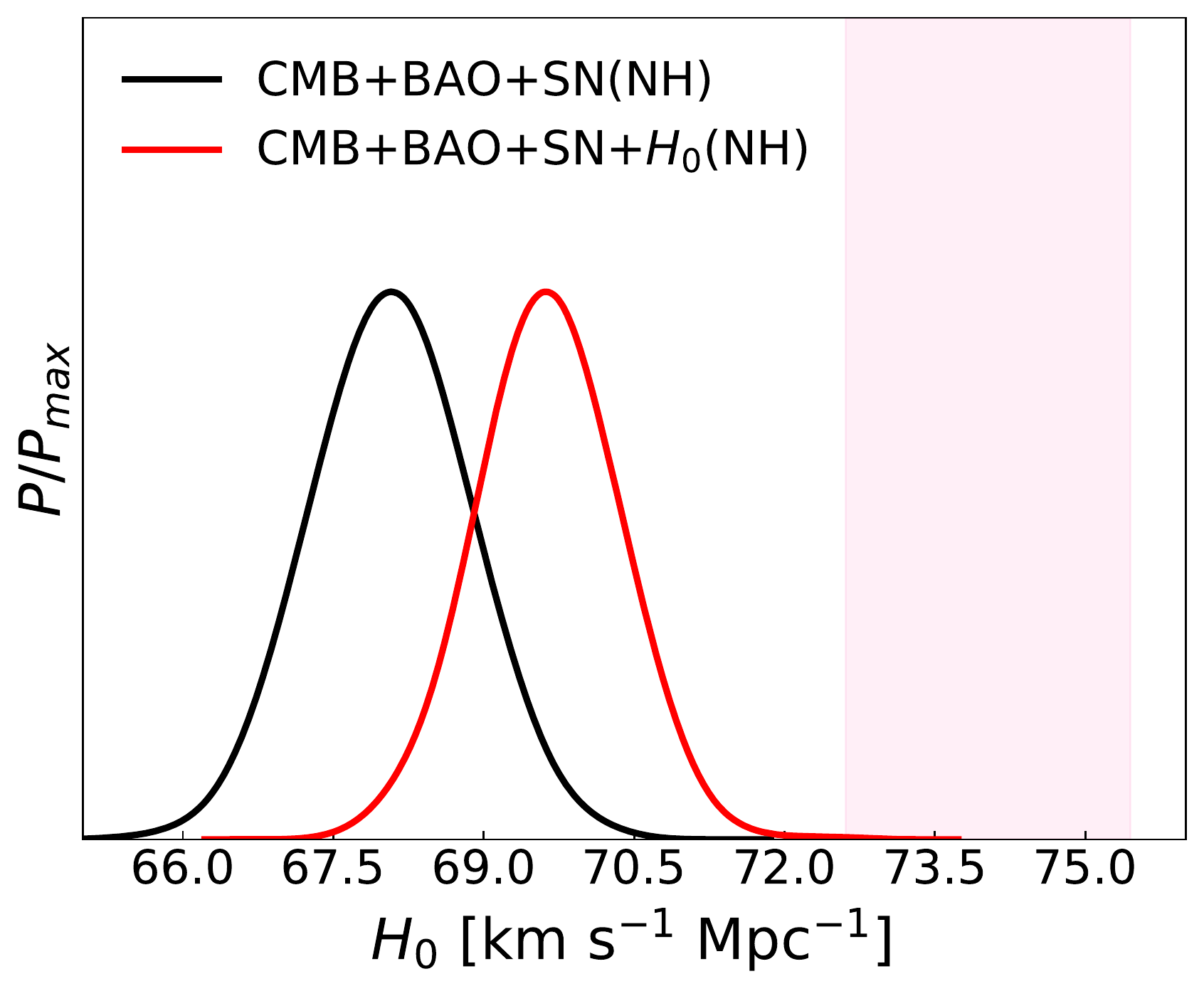}\ \hspace{0.6cm}
\includegraphics[scale=0.3]{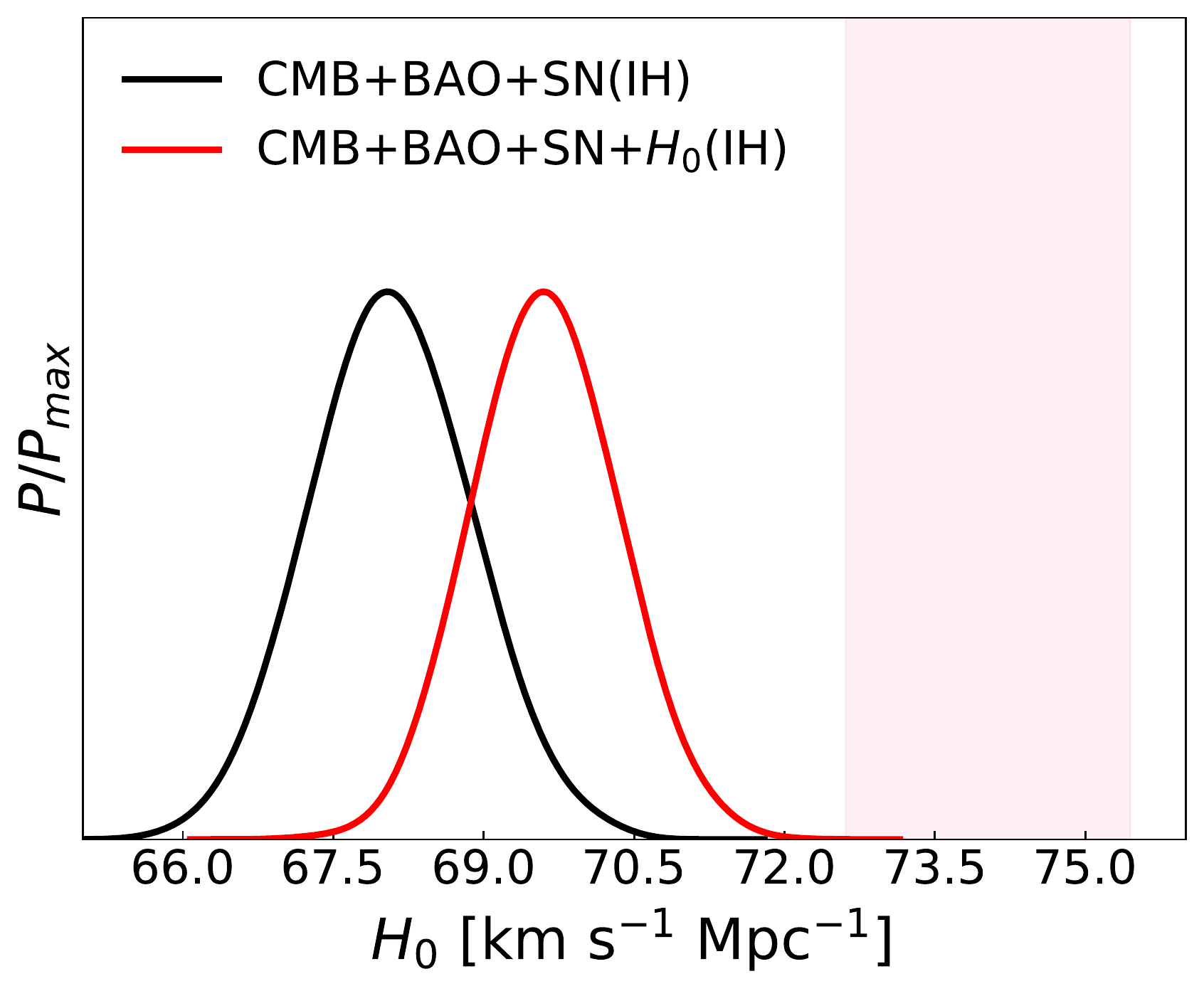}\ \hspace{0.6cm}
\includegraphics[scale=0.3]{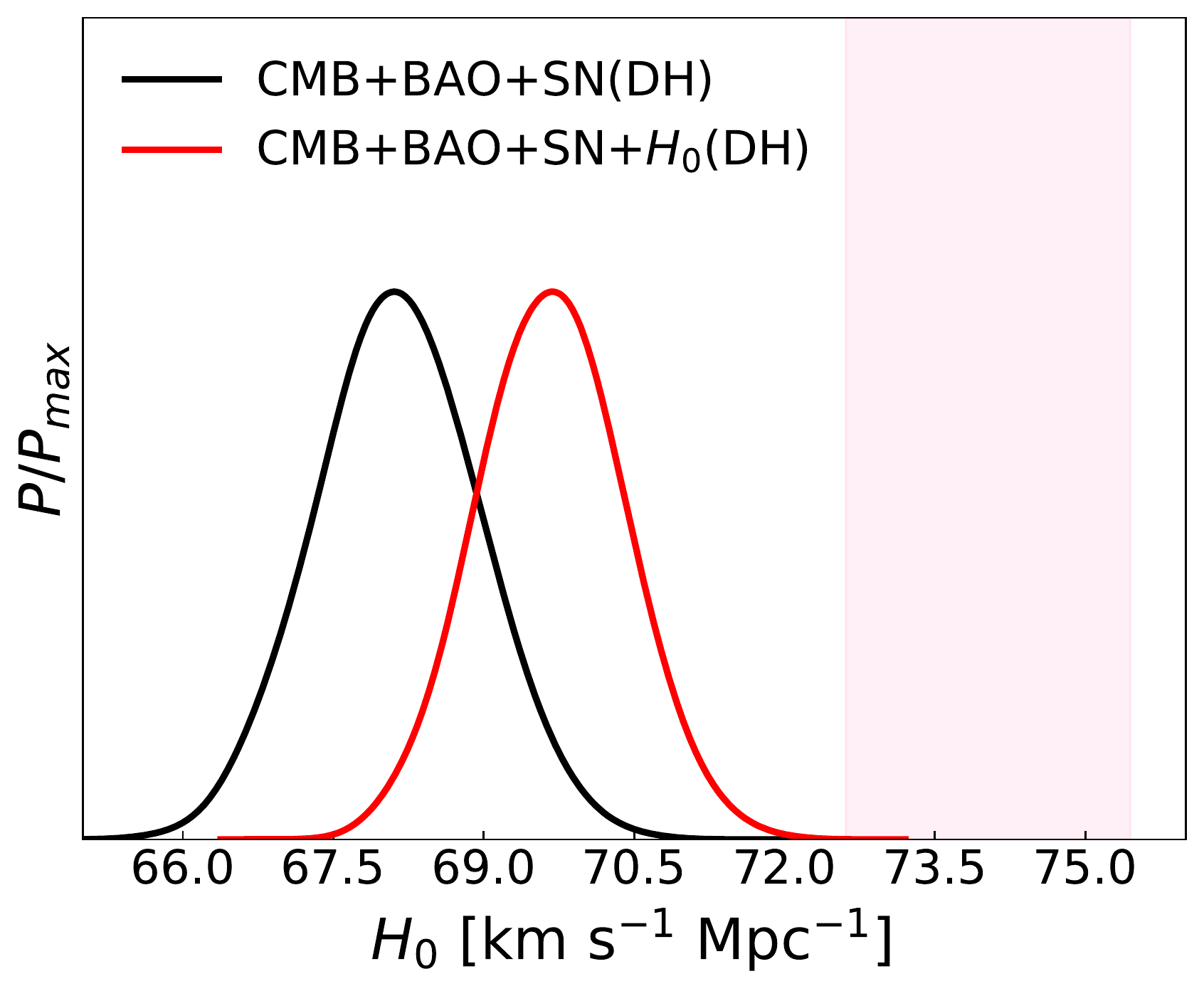}\ \hspace{0.6cm}
\centering \caption{\label{fig7} The one-dimensional posterior distributions for the parameter $H_{0}$ for the I$\Lambda$CDM1+$\sum m_{\nu}$ (NH, IH, and DH) model with $Q=\beta H \rho_{\rm de}$ by using the CMB+BAO+SN and CMB+BAO+SN+$H_{0}$ data combinations, respectively. The result of the local measurement of Hubble constant ($H_{0}=74.03\pm1.42~ \rm{km~ s^{-1} Mpc^{-1}}$) is shown by the hotpink band.}
\end{figure*}

\begin{table*}\small
\setlength\tabcolsep{2pt}
\renewcommand{\arraystretch}{1.5}
\centering
\caption{\label{tab1}Fitting results for the $\Lambda$CDM+$\sum m_{\nu}$ model by using the CMB+BAO+SN and CMB+BAO+SN+$H_{0}$ data combinations, respectively.}
\begin{tabular}{ccccccccccc}

\hline Data &\multicolumn{3}{c}{CMB+BAO+SN}&&\multicolumn{3}{c}{CMB+BAO+SN+$H_{0}$}\\
           \cline{2-4}\cline{6-8}
       Model  & $\Lambda$CDM+$\sum m_{\nu}^{\rm NH}$&$\Lambda$CDM+$\sum m_{\nu}^{\rm IH}$ &$\Lambda$CDM+$\sum m_{\nu}^{\rm DH}$ &&$\Lambda$CDM+$\sum m_{\nu}^{\rm NH}$&$\Lambda$CDM+$\sum m_{\nu}^{\rm IH}$ &$\Lambda$CDM+$\sum m_{\nu}^{\rm DH}$\\

\hline
$\Omega_{\rm m}$                         &$0.3126\pm0.0063$
                                         &$0.3150\pm0.0060$
                                         &$0.3097\pm0.0063$
                                         &&$0.3044\pm0.0056$
                                         &$0.3069\pm0.0056$
                                         &$0.3015\pm0.0056$\\

$H_0\,[{\rm km}/{\rm s}/{\rm Mpc}]$      &$67.48\pm0.47$
                                         &$67.26\pm0.45$
                                         &$67.75\pm0.49$
                                         &&$68.11\pm0.43$
                                         &$67.88\pm0.43$
                                         &$68.40\pm0.44$\\

$\sigma_{8}$                             &$0.801^{+0.011}_{-0.008}$
                                         &$0.793^{+0.010}_{-0.008}$
                                         &$0.812^{+0.013}_{-0.008}$
                                         &&$0.801^{+0.009}_{-0.008}$
                                         &$0.792^{+0.009}_{-0.008}$
                                         &$0.813^{+0.010}_{-0.008}$\\

$\sum m_{\nu}$ [eV]                                      &$<0.156$
                                         &$<0.185$
                                         &$<0.123$
                                         &&$<0.125$
                                         &$<0.160$
                                         &$<0.082$\\

\hline
\end{tabular}

\end{table*}

\begin{table*}\small
\setlength\tabcolsep{2pt}
\renewcommand{\arraystretch}{1.5}
\centering
\caption{\label{tab2}Fitting results for the I$\Lambda$CDM1+$\sum m_{\nu}$ ($Q=\beta H \rho_{\rm de}$) model by using the CMB+BAO+SN and CMB+BAO+SN+$H_{0}$ data combinations, respectively.}
\begin{tabular}{ccccccccccc}

\hline Data &\multicolumn{3}{c}{CMB+BAO+SN}&&\multicolumn{3}{c}{CMB+BAO+SN+$H_{0}$}\\
           \cline{2-4}\cline{6-8}
       Model  & I$\Lambda$CDM+$\sum m_{\nu}^{\rm NH}$&I$\Lambda$CDM+$\sum m_{\nu}^{\rm IH}$ &I$\Lambda$CDM+$\sum m_{\nu}^{\rm DH}$ &&I$\Lambda$CDM+$\sum m_{\nu}^{\rm NH}$&I$\Lambda$CDM+$\sum m_{\nu}^{\rm IH}$ &I$\Lambda$CDM+$\sum m_{\nu}^{\rm DH}$\\

\hline
$\Omega_{\rm m}$                         &$0.285\pm0.029$
                                         &$0.279\pm0.030$
                                         &$0.292^{+0.029}_{-0.030}$
                                         &&$0.235\pm0.027$
                                         &$0.229\pm0.027$
                                         &$0.243^{+0.028}_{-0.027}$\\

$H_0\,[{\rm km}/{\rm s}/{\rm Mpc}]$      &$68.08\pm0.81$
                                         &$68.08^{+0.83}_{-0.82}$
                                         &$68.14\pm0.83$
                                         &&$69.64^{+0.73}_{-0.72}$
                                         &$69.62^{+0.75}_{-0.73}$
                                         &$69.67^{+0.74}_{-0.75}$\\

$\beta$                                  &$0.10^{+0.10}_{-0.11}$
                                         &$0.13\pm0.11$
                                         &$0.07^{+0.10}_{-0.11}$
                                         &&$0.257^{+0.096}_{-0.097}$
                                         &$0.286\pm0.098$
                                         &$0.215\pm0.099$\\

$\sigma_{8}$                             &$0.870^{+0.058}_{-0.088}$
                                         &$0.884^{+0.061}_{-0.094}$
                                         &$0.859^{+0.056}_{-0.085}$
                                         &&$1.013^{+0.076}_{-0.123}$
                                         &$1.031^{+0.081}_{-0.131}$
                                         &$0.987^{+0.073}_{-0.117}$\\

$\sum m_{\nu}$ [eV]                                      &$<0.189$
                                         &$<0.218$
                                         &$<0.151$
                                         &&$<0.177$
                                         &$<0.209$
                                         &$<0.138$\\

\hline
\end{tabular}

\end{table*}
\begin{table*}\small
\setlength\tabcolsep{2pt}
\renewcommand{\arraystretch}{1.5}
\centering
\caption{\label{tab3}Fitting results for the I$\Lambda$CDM2+$\sum m_{\nu}$ ($Q=\beta H \rho_{\rm c}$) model by using the CMB+BAO+SN and CMB+BAO+SN+$H_{0}$ data combinations, respectively.}
\begin{tabular}{ccccccccccc}

\hline Data &\multicolumn{3}{c}{CMB+BAO+SN}&&\multicolumn{3}{c}{CMB+BAO+SN+$H_{0}$}\\
           \cline{2-4}\cline{6-8}
       Model  & I$\Lambda$CDM+$\sum m_{\nu}^{\rm NH}$&I$\Lambda$CDM+$\sum m_{\nu}^{\rm IH}$ &I$\Lambda$CDM+$\sum m_{\nu}^{\rm DH}$ &&I$\Lambda$CDM+$\sum m_{\nu}^{\rm NH}$&I$\Lambda$CDM+$\sum m_{\nu}^{\rm IH}$ &I$\Lambda$CDM+$\sum m_{\nu}^{\rm DH}$\\

\hline
$\Omega_{\rm m}$                         &$0.3085\pm0.0080$
                                         &$0.3092\pm0.0081$
                                         &$0.3077\pm0.0081$
                                         &&$0.2953\pm0.0071$
                                         &$0.2960^{+0.0071}_{-0.0072}$
                                         &$0.2946^{+0.0071}_{-0.0072}$\\

$H_0\,[{\rm km}/{\rm s}/{\rm Mpc}]$      &$67.83\pm0.64$
                                         &$67.74\pm0.65$
                                         &$67.92\pm0.65$
                                         &&$68.92\pm0.60$
                                         &$68.83^{+0.61}_{-0.60}$
                                         &$69.01^{+0.61}_{-0.60}$\\

$\beta$                                  &$0.0011^{+0.0013}_{-0.0012}$
                                         &$0.0014^{+0.0012}_{-0.0013}$
                                         &$0.0005\pm0.0013$
                                         &&$0.0024\pm0.0012$
                                         &$0.0028\pm0.0012$
                                         &$0.0019\pm0.0012$\\

$\sigma_{8}$                             &$0.806^{+0.014}_{-0.012}$
                                         &$0.800^{+0.014}_{-0.012}$
                                         &$0.814^{+0.014}_{-0.013}$
                                         &&$0.815^{+0.013}_{-0.012}$
                                         &$0.809^{+0.013}_{-0.012}$
                                         &$0.824^{+0.014}_{-0.012}$\\

$\sum m_{\nu}$ [eV]                                      &$<0.190$
                                         &$<0.223$
                                         &$<0.149$
                                         &&$<0.170$
                                         &$<0.202$
                                         &$<0.126$\\

\hline
\end{tabular}

\end{table*}
\begin{table*}\small
\setlength\tabcolsep{2pt}
\renewcommand{\arraystretch}{1.5}
\centering
\caption{\label{tab4}Fitting results for the I$\Lambda$CDM3+$\sum m_{\nu}$ ($Q=\beta H_{0} \rho_{\rm de}$) model by using the CMB+BAO+SN and CMB+BAO+SN+$H_{0}$ data combinations, respectively.}
\begin{tabular}{ccccccccccc}

\hline Data &\multicolumn{3}{c}{CMB+BAO+SN}&&\multicolumn{3}{c}{CMB+BAO+SN+$H_{0}$}\\
           \cline{2-4}\cline{6-8}
       Model  & I$\Lambda$CDM+$\sum m_{\nu}^{\rm NH}$&I$\Lambda$CDM+$\sum m_{\nu}^{\rm IH}$ &I$\Lambda$CDM+$\sum m_{\nu}^{\rm DH}$ &&I$\Lambda$CDM+$\sum m_{\nu}^{\rm NH}$&I$\Lambda$CDM+$\sum m_{\nu}^{\rm IH}$ &I$\Lambda$CDM+$\sum m_{\nu}^{\rm DH}$\\

\hline
$\Omega_{\rm m}$                         &$0.282^{+0.036}_{-0.035}$
                                         &$0.276\pm0.035$
                                         &$0.291\pm0.036$
                                         &&$0.223^{+0.032}_{-0.031}$
                                         &$0.217\pm0.033$
                                         &$0.233\pm0.032$\\

$H_0\,[{\rm km}/{\rm s}/{\rm Mpc}]$      &$68.03\pm0.81$
                                         &$67.96^{+0.79}_{-0.80}$
                                         &$68.10\pm0.83$
                                         &&$69.57\pm0.72$
                                         &$69.50\pm0.74$
                                         &$69.63^{+0.73}_{-0.72}$\\

$\beta$                                  &$0.14\pm0.16$
                                         &$0.18\pm0.16$
                                         &$0.08^{+0.16}_{-0.17}$
                                         &&$0.37^{+0.13}_{-0.14}$
                                         &$0.40\pm0.14$
                                         &$0.31\pm0.14$\\

$\sigma_{8}$                             &$0.886^{+0.069}_{-0.118}$
                                         &$0.899^{+0.072}_{-0.122}$
                                         &$0.868^{+0.068}_{-0.113}$
                                         &&$1.072^{+0.095}_{-0.172}$
                                         &$1.10^{+0.10}_{-0.19}$
                                         &$1.036^{+0.091}_{-0.156}$\\

$\sum m_{\nu}$ [eV]                                      &$<0.179$
                                         &$<0.208$
                                         &$<0.140$
                                         &&$<0.166$
                                         &$<0.198$
                                         &$<0.128$\\

\hline
\end{tabular}

\end{table*}
\begin{table*}\small
\setlength\tabcolsep{2pt}
\renewcommand{\arraystretch}{1.5}
\centering
\caption{\label{tab5}Fitting results for the I$\Lambda$CDM4+$\sum m_{\nu}$ ($Q=\beta H_{0} \rho_{\rm c}$) model by using the CMB+BAO+SN and CMB+BAO+SN+$H_{0}$ data combinations, respectively.}
\begin{tabular}{ccccccccccc}

\hline Data &\multicolumn{3}{c}{CMB+BAO+SN}&&\multicolumn{3}{c}{CMB+BAO+SN+$H_{0}$}\\
           \cline{2-4}\cline{6-8}
       Model  & I$\Lambda$CDM+$\sum m_{\nu}^{\rm NH}$&I$\Lambda$CDM+$\sum m_{\nu}^{\rm IH}$ &I$\Lambda$CDM+$\sum m_{\nu}^{\rm DH}$ &&I$\Lambda$CDM+$\sum m_{\nu}^{\rm NH}$&I$\Lambda$CDM+$\sum m_{\nu}^{\rm IH}$ &I$\Lambda$CDM+$\sum m_{\nu}^{\rm DH}$\\

\hline
$\Omega_{\rm m}$                         &$0.299\pm0.016$
                                         &$0.297\pm0.016$
                                         &$0.302^{+0.016}_{-0.017}$
                                         &&$0.272\pm0.013$
                                         &$0.269\pm0.013$
                                         &$0.275\pm0.013$\\

$H_0\,[{\rm km}/{\rm s}/{\rm Mpc}]$      &$68.05\pm0.80$
                                         &$68.07^{+0.83}_{-0.82}$
                                         &$68.07\pm0.81$
                                         &&$69.58\pm0.72$
                                         &$69.58\pm0.73$
                                         &$69.58^{+0.71}_{-0.72}$\\

$\beta$                                  &$0.043\pm0.047$
                                         &$0.058^{+0.047}_{-0.048}$
                                         &$0.024^{+0.047}_{-0.048}$
                                         &&$0.111\pm0.043$
                                         &$0.128\pm0.043$
                                         &$0.092\pm0.044$\\

$\sigma_{8}$                             &$0.814^{+0.019}_{-0.021}$
                                         &$0.812\pm0.019$
                                         &$0.820^{+0.019}_{-0.020}$
                                         &&$0.840\pm0.019$
                                         &$0.837\pm0.019$
                                         &$0.845\pm0.019$\\

$\sum m_{\nu}$ [eV]                                      &$<0.202$
                                         &$<0.235$
                                         &$<0.156$
                                         &&$<0.202$
                                         &$<0.239$
                                         &$<0.162$\\

\hline
\end{tabular}

\end{table*}

\section{Results and discussion}\label{sec3}

In this section, we report the constraint results of cosmological parameters for these I$\Lambda$CDM+$\sum m_{\nu}$ models. The fitting results are listed in Table \ref{tab1} for the $\Lambda$CDM+$\sum m_{\nu}$ model and Tables \ref{tab2}--\ref{tab5} and Figs.~\ref{fig1}--\ref{fig4} for the four I$\Lambda$CDM+$\sum m_{\nu}$ models. For convenience, the I$\Lambda$CDM models with the interaction terms $Q=\beta H \rho_{\rm de}$, $Q=\beta H \rho_{\rm c}$, $Q=\beta H_{0} \rho_{\rm de}$ and $Q=\beta H_{0} \rho_{\rm c}$ are denoted as ``I$\Lambda$CDM1", ``I$\Lambda$CDM2", ``I$\Lambda$CDM3", and ``I$\Lambda$CDM4", respectively. In these tables, we show the best fit values with $\pm 1 \sigma$ errors of the cosmological parameters, but for the total neutrino mass $\sum m_{\nu}$, which cannot be well constrained, the $2\sigma$ upper limits are given.

\subsection{Neutrino mass}\label{seca}

Firstly, we use the CMB+BAO+SN data combination to constrain these models. In the $\Lambda$CDM+$\sum m_{\nu}$ model, we obtain $\sum m_{\nu}<0.156$ eV for the NH case, $\sum m_{\nu}<0.185$ eV for the IH case, and $\sum m_{\nu}<0.123$ eV for the DH case, as shown Table \ref{tab1}. In the I$\Lambda$CDM1+$\sum m_{\nu}$ model, the constraint results are $\sum m_{\nu}<0.187$ eV for the NH case, $\sum m_{\nu}<0.218$ eV for the IH case, and $\sum m_{\nu}<0.151$ eV for the DH case (see Table \ref{tab2}); in the I$\Lambda$CDM2+$\sum m_{\nu}$ model, the results are $\sum m_{\nu}<0.190$ eV for the NH case, $\sum m_{\nu}<0.223$ eV for the IH case, and $\sum m_{\nu}<0.149$ eV for the DH case (see Table \ref{tab3}); in the I$\Lambda$CDM3+$\sum m_{\nu}$ model, we get $\sum m_{\nu}<0.179$ eV for the NH case, $\sum m_{\nu}<0.208$ eV for the IH case, and $\sum m_{\nu}<0.140$ eV for the DH case (see Table \ref{tab4}); in the I$\Lambda$CDM4+$\sum m_{\nu}$ model, the constraint results become $\sum m_{\nu}<0.202$ eV for the NH case, $\sum m_{\nu}<0.235$ eV for the IH case, and $\sum m_{\nu}<0.156$ eV for the DH case (see Table \ref{tab5}). We find that, the constraint results of $\sum m_{\nu}$ are looser in the four I$\Lambda$CDM+$\sum m_{\nu}$ models than those in the $\Lambda$CDM+$\sum m_{\nu}$ model. When considering the three mass hierarchies, we find that the constraint results of $\sum m_{\nu}$ are tightest in the DH case and loosest in the IH case (see the left panels in Figs.~\ref{fig1}--\ref{fig4}).

Then, we consider the data combination involving the latest local measurement of the Hubble constant $H_{0}$ to constrain these models. By using the CMB+BAO+SN+$H_{0}$ data combination, we find that the constraint results of $\sum m_{\nu}$ are looser in the four I$\Lambda$CDM+$\sum m_{\nu}$ models than those in the $\Lambda$CDM+$\sum m_{\nu}$ model, and when considering the three mass hierarchies, the constraint results of $\sum m_{\nu}$ are tightest in the DH case and loosest in the IH case. These conclusions are consistent with the case using the CMB+BAO+SN data combination. Additionally, we also find that the constraints on $\sum m_{\nu}$ become slightly tighter for using CMB+BAO+SN+$H_{0}$ than CMB+BAO+SN.

\subsection{Coupling parameter}\label{secb}

In this subsection, we discuss the fitting results of the coupling parameter $\beta$ in these four I$\Lambda$CDM+$\sum m_{\nu}$ models by using the CMB+BAO+SN and CMB+BAO+SN+$H_{0}$ data combinations, respectively.

First, we constrain the I$\Lambda$CDM1+$\sum m_{\nu}$ model (see Table \ref{tab2}) using the CMB+BAO+SN data combination, and we obtain $\beta=0.10^{+0.10}_{-0.11}$ for the NH case, $\beta=0.13\pm0.11$ for the IH case, and $\beta=0.07^{+0.10}_{-0.11}$ for the DH case. It is shown that a positive value of $\beta$ is favored and $\beta>0$ is at the $0.91\sigma$, $1.18\sigma$, and $0.64\sigma$ levels for the three mass hierarchy cases, respectively. Furthermore, we constrain this model by using the CMB+BAO+SN+$H_{0}$ data combination, and we obtain $\beta=0.257^{+0.096}_{-0.097}$ for the NH case, $\beta=0.286\pm0.098$ for the IH case, and $\beta=0.215\pm0.099$ for the DH case. Now, $\beta>0$ is obtained at the $2.65\sigma$, $2.92\sigma$, and $2.17\sigma$ levels, respectively. This indicates that cold dark matter decaying into dark energy is favored when using the CMB+BAO+SN+$H_{0}$ data combination.

For the I$\Lambda$CDM2+$\sum m_{\nu}$ model (see Table \ref{tab3}), we obtain $\beta=0.0011^{+0.0013}_{-0.0012}$ for the NH case, $\beta=0.0014^{+0.0012}_{-0.0013}$ for the IH case, and $\beta=0.0005\pm0.0013$ for the DH case, by using the CMB+BAO+SN data combination. Thus, $\beta>0$ is favored at the $0.92\sigma$, $1.08\sigma$, and $0.38\sigma$ levels, respectively. When using the CMB+BAO+SN+$H_{0}$ data combination, we obtain $\beta=0.0024\pm0.0012$ for the NH case, $\beta=0.0028\pm0.0012$ for the IH case, and $\beta=0.0019\pm0.0012$ for the DH case, which indicates that a positive value of $\beta$ can be detected at the $2.00\sigma$, $2.33\sigma$, and $1.58\sigma$ levels, respectively.

As for the I$\Lambda$CDM3+$\sum m_{\nu}$ model (see Table \ref{tab4}), we obtain $\beta=0.14\pm0.16$ for the NH case, $\beta=0.18\pm0.16$ for the IH case, and $\beta=0.08^{+0.16}_{-0.17}$ for the DH case, by using the CMB+BAO+SN data combination. Therefore, the positive values of $\beta$ are favored and $\beta>0$ is preferred at the $0.88\sigma$, $1.13\sigma$, and $0.50\sigma$ levels, respectively. When using the CMB+BAO+SN+$H_{0}$ data combination, we obtain $\beta=0.37^{+0.13}_{-0.14}$ for the NH case, $\beta=0.40\pm0.14$ for the IH case, and $\beta=0.31\pm0.14$ for the DH case, respectively. And $\beta>0$ is detected at the $2.64\sigma$, $2.90\sigma$, and $2.21\sigma$ levels, respectively, which indicates cold dark matter decaying into dark energy.

Finally, we show the constraint results of I$\Lambda$CDM4+$\sum m_{\nu}$ model (see Table \ref{tab5}). We obtain $\beta=0.043\pm0.047$ for the NH case, $\beta=0.058^{+0.047}_{-0.048}$ for the IH case, and $\beta=0.024^{+0.047}_{-0.048}$ for the DH case, by using the CMB+BAO+SN data combination. So, a positive value of $\beta$ is favored and $\beta>0$ is at the $0.91\sigma$, $1.21\sigma$, and $0.50\sigma$ levels, respectively. When using the CMB+BAO+SN+$H_{0}$ data combination, we obtain $\beta=0.111\pm0.043$ for the NH case, $\beta=0.128\pm0.043$ for the IH case, and $\beta=0.092\pm0.044$ for the DH case. Now, $\beta>0$ is preferred at the $2.58\sigma$, $2.98\sigma$, and $2.09\sigma$ levels, respectively. The conclusion is the same as the above three cases, i.e., cold dark matter decaying into dark energy is supported by the CMB+BAO+SN+$H_{0}$ data combination.

In summary, for all the I$\Lambda$CDM+$\sum m_{\nu}$ models considered in this paper, the values of $\beta$ are greater by using the CMB+BAO+SN+$H_{0}$ data combination than using CMB+BAO+SN data combination. We can also intuitively obtain this conclusion by comparing the left and right panels of Figs.~\ref{fig1}--\ref{fig4}. Additionally, when using CMB+BAO+SN+$H_{0}$ data combination, $\beta>0$ is favored at more than 1$\sigma$ level in all the I$\Lambda$CDM+$\sum m_{\nu}$ models, which indicates that cold dark matter decaying into dark energy is supported in these models.

\subsection{The $H_{0}$ tension}\label{secc}

In this subsection, we discuss the issue of $H_{0}$ tension between the Planck observation of the CMB power spectra and the local measurement based on the method of distance ladder. The detailed fitting results are given in Tables \ref{tab1}--\ref{tab5} and Figs.~\ref{fig5}--\ref{fig7}.

In Table \ref{tab1}, we show the constraint results of the $\Lambda$CDM+$\sum m_{\nu}$ model by using the CMB+BAO+SN data combination. We obtain $H_{0}=67.48\pm0.47~ \rm{km~ s^{-1} Mpc^{-1}}$ for the NH case, $H_{0}=67.26\pm0.45~ \rm{km~ s^{-1} Mpc^{-1}}$ for the IH case, and $H_{0}=67.75\pm0.49~ \rm{km~ s^{-1} Mpc^{-1}}$ for the DH case, which are $4.38 \sigma$ $4.54 \sigma$ and $4.18 \sigma$ lower than the direct measurement of the Hubble constant ($H_{0}=74.03\pm1.42~ \rm{km~ s^{-1} Mpc^{-1}}$). So, we investigate whether the $H_{0}$ tension can be solved or relieved in the interacting dark energy scenario. In Tables \ref{tab2}--\ref{tab5}, we show the constraint results of the I$\Lambda$CDM1+$\sum m_{\nu}$, I$\Lambda$CDM2+$\sum m_{\nu}$, I$\Lambda$CDM3+$\sum m_{\nu}$, and I$\Lambda$CDM4+$\sum m_{\nu}$ models from the CMB+BAO+SN data combination. In the I$\Lambda$CDM1+$\sum m_{\nu}$ model, we obtain $H_{0}=68.08\pm0.81~ \rm{km~ s^{-1} Mpc^{-1}}$ for the NH case, $H_{0}=68.08^{+0.83}_{-0.82}~ \rm{km~ s^{-1} Mpc^{-1}}$ for the IH case, and $H_{0}=68.14\pm0.83~ \rm{km~ s^{-1} Mpc^{-1}}$ for the DH case; in the I$\Lambda$CDM2+$\sum m_{\nu}$ model, we obtain $H_{0}=67.83\pm0.64~ \rm{km~ s^{-1} Mpc^{-1}}$ for the NH case, $H_{0}=67.74\pm0.65~ \rm{km~ s^{-1} Mpc^{-1}}$ for the IH case, and $H_{0}=67.92\pm0.65~ \rm{km~ s^{-1} Mpc^{-1}}$ for the DH case; in the I$\Lambda$CDM3+$\sum m_{\nu}$ model, we obtain $H_{0}=68.03\pm0.81~ \rm{km~ s^{-1} Mpc^{-1}}$ for the NH case, $H_{0}=67.96^{+0.79}_{-0.80}~ \rm{km~ s^{-1} Mpc^{-1}}$ for the IH case, and $H_{0}=68.10\pm0.83~ \rm{km~ s^{-1} Mpc^{-1}}$ for the DH case; in the I$\Lambda$CDM4+$\sum m_{\nu}$ model, we obtain $H_{0}=68.05\pm0.80~ \rm{km~ s^{-1} Mpc^{-1}}$ for the NH case, $H_{0}=68.07^{+0.83}_{-0.82}~ \rm{km~ s^{-1} Mpc^{-1}}$ for the IH case, and $H_{0}=68.07\pm0.81~ \rm{km~ s^{-1} Mpc^{-1}}$ for the DH case. For these cases, the tensions with the Hubble constant direct measurement are at the $3.64 \sigma$ level, $3.62 \sigma$ level, $3.58 \sigma$ level, $3.98 \sigma$ level, $4.03 \sigma$ level, $3.91 \sigma$ level, $3.67 \sigma$ level, $3.74 \sigma$ level, $3.61 \sigma$ level, $3.67 \sigma$ level, $3.62 \sigma$ level, and $3.65 \sigma$ level, respectively.

Then, we show the constraint results of these models by using the CMB+BAO+SN+$H_{0}$ data combination (see Tables \ref{tab1}--\ref{tab5}). In the $\Lambda$CDM+$\sum m_{\nu}$ model, we obtain $H_{0}=68.11\pm0.43~ \rm{km~ s^{-1} Mpc^{-1}}$ for the NH case, $H_{0}=67.88\pm0.43~ \rm{km~ s^{-1} Mpc^{-1}}$ for the IH case, and $H_{0}=68.40\pm0.44~ \rm{km~ s^{-1} Mpc^{-1}}$ for the DH case, which indicates that the tensions with the Hubble constant direct measurement are at the $3.99 \sigma$ level $4.14 \sigma$ level and $3.79 \sigma$ level, respectively. In the I$\Lambda$CDM1+$\sum m_{\nu}$ model, we obtain $H_{0}=69.64^{+0.73}_{-0.72}~ \rm{km~ s^{-1} Mpc^{-1}}$ for the NH case, $H_{0}=69.62^{+0.75}_{-0.73}~ \rm{km~ s^{-1} Mpc^{-1}}$ for the IH case, and $H_{0}=69.67^{+0.74}_{-0.75}~ \rm{km~ s^{-1} Mpc^{-1}}$ for the DH case; in the I$\Lambda$CDM2+$\sum m_{\nu}$ model, we obtain $H_{0}=68.92\pm0.60~ \rm{km~ s^{-1} Mpc^{-1}}$ for the NH case, $H_{0}=68.83^{+0.61}_{-0.60}~ \rm{km~ s^{-1} Mpc^{-1}}$ for the IH case, and $H_{0}=69.01^{+0.61}_{-0.60}~ \rm{km~ s^{-1} Mpc^{-1}}$ for the DH case; in the I$\Lambda$CDM3+$\sum m_{\nu}$ model, we obtain $H_{0}=69.57\pm0.72~ \rm{km~ s^{-1} Mpc^{-1}}$ for the NH case, $H_{0}=69.50\pm0.74~ \rm{km~ s^{-1} Mpc^{-1}}$ for the IH case, and $H_{0}=69.63^{+0.73}_{-0.72}~ \rm{km~ s^{-1} Mpc^{-1}}$ for the DH case; in the I$\Lambda$CDM4+$\sum m_{\nu}$ model, we obtain $H_{0}=69.58\pm0.72~ \rm{km~ s^{-1} Mpc^{-1}}$ for the NH case, $H_{0}=69.58\pm0.73~ \rm{km~ s^{-1} Mpc^{-1}}$ for the IH case, and $H_{0}=69.58^{+0.71}_{-0.72}~ \rm{km~ s^{-1} Mpc^{-1}}$ for the DH case. The tensions with the Hubble constant direct measurement are at the $2.75 \sigma$ level, $2.75 \sigma$ level, $2.72 \sigma$ level, $3.31 \sigma$ level, $3.36 \sigma$ level, $3.25 \sigma$ level, $2.80 \sigma$ level, $2.83 \sigma$ level, $2.76 \sigma$ level, $2.80 \sigma$ level, $2.79 \sigma$ level, and $2.80 \sigma$ level, respectively.

From the above constraint results, we find that compared with the $\Lambda$CDM+$\sum m_{\nu}$ model, the $H_{0}$ tension can indeed be relieved in the I$\Lambda$CDM+$\sum m_{\nu}$ model. From Figs.~\ref{fig5}--\ref{fig6} we can clearly see that for whichever neutrino mass hierarchy case, the fitting values of $H_{0}$ in the I$\Lambda$CDM+$\sum m_{\nu}$ models (here, we take I$\Lambda$CDM1+$\sum m_{\nu}$ with $Q=\beta H \rho_{\rm de}$ as an example) are always much larger than those in the $\Lambda$CDM+$\sum m_{\nu}$ model. We also find that, the CMB+BAO+SN+$H_{0}$ data combination (about $2.7-3.4\sigma$ level) is slightly more effective in relieving the $H_{0}$ tension than the CMB+BAO+SN data combination (about $3.6-4.0 \sigma$ level), due to the employment of the $H_{0}$ prior in the data combination. To visually display the result, we also take the I$\Lambda$CDM1+$\sum m_{\nu}$ model as an example to give this result in Fig.~\ref{fig7}. From these figures, we can clearly see that for whichever hierarchy of the neutrino mass spectrum, the values of $H_{0}$ are always much larger when adding the $H_{0}$ data in a cosmological fit. Certainly, the $H_{0}$ tension problem only can be alleviated to some extent in these cases, but cannot be truly solved. For the issue of $H_{0}$ tension, further exploration is needed.

\section{Conclusion}\label{sec4}

In this work, we have investigated the constraints on the total neutrino mass in the scenario of vacuum energy interacting with cold dark matter by using the latest cosmological observations. We consider four typical models, i.e., I$\Lambda$CDM1+$\sum m_{\nu}$ ($Q=\beta H \rho_{\rm de}$) model, I$\Lambda$CDM2+$\sum m_{\nu}$ ($Q=\beta H \rho_{\rm c}$) model, I$\Lambda$CDM3+$\sum m_{\nu}$ ($Q=\beta H_{0} \rho_{\rm de}$) model, and I$\Lambda$CDM4+$\sum m_{\nu}$ ($Q=\beta H_{0} \rho_{\rm c}$) model. For the three-generation neutrinos, we consider the NH, IH, and DH cases. We employ the extended version of the parameterized post-Friedmann approach to calculate the perturbation of dark energy in the IDE cosmology. We use the Planck 2018 CMB data, the BAO measurements, the SN data of Pantheon compilation, and the local measurement of the Hubble constant $H_{0}$ from the Hubble Space Telescope to constrain these models.

We find that, compared with the $\Lambda$CDM+$\sum m_{\nu}$ model, these four I$\Lambda$CDM+$\sum m_{\nu}$ models can provide a much looser constraint on the total neutrino mass $\sum m_{\nu}$. When considering the three mass hierarchies, the upper limits on $\sum m_{\nu}$ are smallest in the DH case and largest in the IH case. We also find that, the constraints on $\sum m_{\nu}$ are slightly tighter by using CMB+BAO+SN+$H_{0}$ than using CMB+BAO+SN. In addition, in all the I$\Lambda$CDM+$\sum m_{\nu}$ models considered in this paper, the fit values of $\beta$ are greater using the CMB+BAO+SN+$H_{0}$ data combination than using the CMB+BAO+SN data combination, and $\beta>0$ is favored at more than 1$\sigma$ level in all the I$\Lambda$CDM+$\sum m_{\nu}$ models when using the CMB+BAO+SN+$H_{0}$ data combination, implying the preference of cold dark matter decaying into dark energy. In addition, we also find that, compared with the $\Lambda$CDM+$\sum m_{\nu}$ model, the $H_{0}$ tension can be alleviated to some extent in the I$\Lambda$CDM+$\sum m_{\nu}$ models.

\begin{acknowledgments}
This work was supported by the National Natural Science Foundation of China (Grant Nos. 11975072, 11875102, 11835009, and 11690021), the Liaoning Revitalization Talents Program (Grant No. XLYC1905011), the Fundamental Research Funds for the Central Universities (Grant No. N2005030), and the Top-Notch Young Talents Program of China (W02070050).

\end{acknowledgments}


\begin{thebibliography}{99}



\bibitem{Lesgourgues:2006nd}
  J.~Lesgourgues and S.~Pastor,
  Massive neutrinos and cosmology,
  Phys.\ Rept.\  {\bf 429}, 307 (2006)
  doi:10.1016/j.physrep.2006.04.001
  [astro-ph/0603494].

\bibitem{Agashe:2014kda}
  K.~A.~Olive {\it et al.} [Particle Data Group],
  Review of Particle Physics,
  Chin.\ Phys.\ C {\bf 38}, 090001 (2014).
  doi:10.1088/1674-1137/38/9/090001

\bibitem{Xing:2019vks}
  Z.~z.~Xing,
  Flavor structures of charged fermions and massive neutrinos,
  Phys.\ Rept.\  {\bf 854}, 1 (2020)
  doi:10.1016/j.physrep.2020.02.001
  [arXiv:1909.09610 [hep-ph]].

\bibitem{Osipowicz:2001sq}
  A.~Osipowicz {\it et al.} [KATRIN Collaboration],
  KATRIN: A Next generation tritium beta decay experiment with sub-eV sensitivity for the electron neutrino mass. Letter of intent,
  hep-ex/0109033.

\bibitem{KlapdorKleingrothaus:2002ip}
  H.~V.~Klapdor-Kleingrothaus and U.~Sarkar,
  Implications of observed neutrinoless double beta decay,
  Mod.\ Phys.\ Lett.\ A {\bf 16}, 2469 (2001)
  doi:10.1142/S0217732301005850

\bibitem{KlapdorKleingrothaus:2004wj}
  H.~V.~Klapdor-Kleingrothaus, I.~V.~Krivosheina, A.~Dietz and O.~Chkvorets,
  Search for neutrinoless double beta decay with enriched Ge-76 in Gran Sasso 1990-2003,
  Phys.\ Lett.\ B {\bf 586}, 198 (2004)
  doi:10.1016/j.physletb.2004.02.025
  [hep-ph/0404088].

\bibitem{Kraus:2004zw}
  C.~Kraus {\it et al.},
  Final results from phase II of the Mainz neutrino mass search in tritium beta decay,
  Eur.\ Phys.\ J.\ C {\bf 40}, 447 (2005)
  doi:10.1140/epjc/s2005-02139-7
  [hep-ex/0412056].

\bibitem{Otten:2008zz}
  E.~W.~Otten and C.~Weinheimer,
  Neutrino mass limit from tritium beta decay,
  Rept.\ Prog.\ Phys.\  {\bf 71}, 086201 (2008)
  doi:10.1088/0034-4885/71/8/086201
  [arXiv:0909.2104 [hep-ex]].

\bibitem{Wolf:2008hf}
  J.~Wolf [KATRIN Collaboration],
  The KATRIN Neutrino Mass Experiment,
  Nucl.\ Instrum.\ Meth.\ A {\bf 623}, 442 (2010)
  doi:10.1016/j.nima.2010.03.030
  [arXiv:0810.3281 [physics.ins-det]].

\bibitem{Huang:2016qmh}
  G.~y.~Huang and S.~Zhou,
  Discriminating between thermal and nonthermal cosmic relic neutrinos through an annual modulation at PTOLEMY,
  Phys.\ Rev.\ D {\bf 94}, no. 11, 116009 (2016)
  doi:10.1103/PhysRevD.94.116009
  [arXiv:1610.01347 [hep-ph]].

\bibitem{Zhang:2017ljh}
  J.~Zhang and X.~Zhang,
  Gravitational clustering of cosmic relic neutrinos in the Milky Way,
  Nature Commun.\  {\bf 9}, 1833 (2018)
  doi:10.1038/s41467-018-04264-y
  [arXiv:1712.01153 [astro-ph.CO]].

\bibitem{Betts:2013uya}
  S.~Betts {\it et al.},
  Development of a Relic Neutrino Detection Experiment at PTOLEMY: Princeton Tritium Observatory for Light, Early-Universe, Massive-Neutrino Yield,
  arXiv:1307.4738 [astro-ph.IM].

\bibitem{Valle:2005ai}
  J.~W.~F.~Valle,
  Neutrino masses and oscillations,
  AIP Conf.\ Proc.\  {\bf 805}, no. 1, 128 (2005)
  doi:10.1063/1.2149688
  [hep-ph/0509262].

\bibitem{Hannestad:2010kz}
  S.~Hannestad,
  Neutrino physics from precision cosmology,
  Prog.\ Part.\ Nucl.\ Phys.\  {\bf 65}, 185 (2010)
  doi:10.1016/j.ppnp.2010.07.001
  [arXiv:1007.0658 [hep-ph]].

\bibitem{Lesgourgues:2012uu}
  J.~Lesgourgues and S.~Pastor,
  Neutrino mass from Cosmology,
  Adv.\ High Energy Phys.\  {\bf 2012}, 608515 (2012)
  doi:10.1155/2012/608515
  [arXiv:1212.6154 [hep-ph]].

\bibitem{Abazajian:2013oma}
  K.~N.~Abazajian {\it et al.} [Topical Conveners: K.N. Abazajian, J.E. Carlstrom, A.T. Lee Collaboration],
  Neutrino Physics from the Cosmic Microwave Background and Large Scale Structure,
  Astropart.\ Phys.\  {\bf 63}, 66 (2015)
  doi:10.1016/j.astropartphys.2014.05.014
  [arXiv:1309.5383 [astro-ph.CO]].

\bibitem{Zhang:2020mox}
  M.~Zhang, J.~F.~Zhang and X.~Zhang,
  Impacts of dark energy on constraining neutrino mass after Planck 2018,
  arXiv:2005.04647 [astro-ph.CO].

\bibitem{Hu:1997mj}
  W.~Hu, D.~J.~Eisenstein and M.~Tegmark,
  Weighing neutrinos with galaxy surveys,
  Phys.\ Rev.\ Lett.\  {\bf 80}, 5255 (1998).

\bibitem{Reid:2009nq}
  B.~A.~Reid, L.~Verde, R.~Jimenez and O.~Mena,
  Robust neutrino constraints by combining low redshift observations with the CMB,
  JCAP {\bf 1001}, 003 (2010).

\bibitem{Thomas:2009ae}
  S.~A.~Thomas, F.~B.~Abdalla and O.~Lahav,
  Upper bound of 0.28eV on the neutrino masses from the largest photometric redshift survey,
  Phys.\ Rev.\ Lett.\  {\bf 105}, 031301 (2010).

\bibitem{Carbone:2010ik}
  C.~Carbone, L.~Verde, Y.~Wang and A.~Cimatti,
  Neutrino constraints from future nearly all-sky spectroscopic galaxy surveys,
  JCAP {\bf 1103}, 030 (2011).


\bibitem{Li:2012vn}
  H.~Li and X.~Zhang,
  Constraining dynamical dark energy with a divergence-free parametrization in the presence of spatial curvature and massive neutrinos,
  Phys.\ Lett.\ B {\bf 713}, 160 (2012).

\bibitem{Wang:2012vh}
  X.~Wang, X.~L.~Meng, T.~J.~Zhang, H.~Shan, Y.~Gong, C.~Tao, X.~Chen and Y.~F.~Huang,
  Observational constraints on cosmic neutrinos and dark energy revisited,
  JCAP {\bf 1211}, 018 (2012)
  doi:10.1088/1475-7516/2012/11/018
  [arXiv:1210.2136 [astro-ph.CO]].

\bibitem{Wang:2012uf}
  Y.~H.~Li, S.~Wang, X.~D.~Li and X.~Zhang,
  Holographic dark energy in a universe with spatial curvature and massive neutrinos: a full Markov Chain Monte Carlo exploration,
  JCAP {\bf 1302}, 033 (2013).


\bibitem{Audren:2012vy}
  B.~Audren, J.~Lesgourgues, S.~Bird, M.~G.~Haehnelt and M.~Viel,
  Neutrino masses and cosmological parameters from a Euclid-like survey: Markov Chain Monte Carlo forecasts including theoretical errors,
  JCAP {\bf 1301}, 026 (2013).

\bibitem{Riemer-Sorensen:2013jsa}
  S.~Riemer-S$\phi$rensen, D.~Parkinson and T.~M.~Davis,
  Combining Planck data with large scale structure information gives a strong neutrino mass constraint,
  Phys.\ Rev.\ D {\bf 89}, 103505 (2014).

\bibitem{Font-Ribera:2013rwa}
  A.~Font-Ribera, P.~McDonald, N.~Mostek, B.~A.~Reid, H.~J.~Seo and A.~Slosar,
  DESI and other dark energy experiments in the era of neutrino mass measurements,
  JCAP {\bf 1405}, 023 (2014).



\bibitem{Zhang:2014dxk}
  J.~F.~Zhang, Y.~H.~Li and X.~Zhang,
  Sterile neutrinos help reconcile the observational results of primordial gravitational waves from Planck and BICEP2,
  Phys.\ Lett.\ B {\bf 740}, 359 (2015).


\bibitem{Zhang:2014nta}
  J.~F.~Zhang, Y.~H.~Li and X.~Zhang,
  Cosmological constraints on neutrinos after BICEP2,
  Eur.\ Phys.\ J.\ C {\bf 74}, 2954 (2014).


\bibitem{Zhang:2014ifa}
  J.~F.~Zhang, J.~J.~Geng and X.~Zhang,
  Neutrinos and dark energy after Planck and BICEP2: data consistency tests and cosmological parameter constraints,
  JCAP {\bf 1410}, no. 10, 044 (2014).



\bibitem{Palanque-Delabrouille:2014jca}
  N.~Palanque-Delabrouille {\it et al.},
  Constraint on neutrino masses from SDSS-III/BOSS Ly$\alpha$ forest and other cosmological probes,
  JCAP {\bf 1502}, no. 02, 045 (2015).

\bibitem{Geng:2014yoa}
  C.~Q.~Geng, C.~C.~Lee and J.~L.~Shen,
  Matter power spectra in viable $f(R)$ gravity models with massive neutrinos,
  Phys.\ Lett.\ B {\bf 740}, 285 (2015).

\bibitem{Li:2015poa}
  Y.~H.~Li, J.~F.~Zhang and X.~Zhang,
  Probing $f(R)$ cosmology with sterile neutrinos via measurements of scale-dependent growth rate of structure,
  Phys.\ Lett.\ B {\bf 744}, 213 (2015).


\bibitem{Ade:2015xua}
   P.~A.~R.~Ade {\it et al.} [Planck Collaboration],
  Planck 2015 results. XIII. cosmological parameters,
   Astron.\ Astrophys.\  {\bf 594}, A13 (2016).

\bibitem{Zhang:2015rha}
  J.~F.~Zhang, M.~M.~Zhao, Y.~H.~Li and X.~Zhang,
 Neutrinos in the holographic dark energy model: constraints from latest measurements of expansion history and growth of structure,
  JCAP {\bf 1504}, 038 (2015).

\bibitem{Geng:2015haa}
  C.~Q.~Geng, C.~C.~Lee, R.~Myrzakulov, M.~Sami and E.~N.~Saridakis,
 Observational constraints on varying neutrino-mass cosmology,
  JCAP {\bf 1601}, no. 01, 049 (2016).

\bibitem{Chen:2015oga}
  Y.~Chen and L.~Xu,
 Galaxy clustering, CMB and supernova data constraints on $\varphi$CDM model with massive neutrinos,
  Phys.\ Lett.\ B {\bf 752}, 66 (2016).




\bibitem{Allison:2015qca}
  R.~Allison, P.~Caucal, E.~Calabrese, J.~Dunkley and T.~Louis,
  Towards a cosmological neutrino mass detection,
  Phys.\ Rev.\ D {\bf 92}, no. 12, 123535 (2015).

\bibitem{Cuesta:2015iho}
  A.~J.~Cuesta, V.~Niro and L.~Verde,
 Neutrino mass limits: robust information from the power spectrum of galaxy surveys,
  Phys.\ Dark Univ.\  {\bf 13}, 77 (2016).





\bibitem{Chen:2016eyp}
  Y.~Chen, B.~Ratra, M.~Biesiada, S.~Li and Z.~H.~Zhu,
 Constraints on non-flat cosmologies with massive neutrinos after Planck 2015,
  Astrophys.\ J.\  {\bf 829}, no. 2, 61 (2016).

\bibitem{Moresco:2016nqq}
  M.~Moresco, R.~Jimenez, L.~Verde, A.~Cimatti, L.~Pozzetti, C.~Maraston and D.~Thomas,
  Constraining the time evolution of dark energy, curvature and neutrino properties with cosmic chronometers,
  JCAP {\bf 1612}, no. 12, 039 (2016).


\bibitem{Lu:2016hsd}
  J.~Lu, M.~Liu, Y.~Wu, Y.~Wang and W.~Yang,
  Cosmic constraint on massive neutrinos in viable $f(R)$ gravity with producing $\Lambda$CDM background expansion,
  Eur.\ Phys.\ J.\ C {\bf 76}, no. 12, 679 (2016).

\bibitem{Kumar:2016zpg}
  S.~Kumar and R.~C.~Nunes,
  Probing the interaction between dark matter and dark energy in the presence of massive neutrinos,
  Phys.\ Rev.\ D {\bf 94}, no. 12, 123511 (2016).

\bibitem{Xu:2016ddc}
  L.~Xu and Q.~G.~Huang,
  Detecting the neutrinos mass hierarchy from cosmological data,
  Sci.\ China Phys.\ Mech.\ Astron.\  {\bf 61}, no. 3, 039521 (2018).




\bibitem{Vagnozzi:2017ovm}
  S.~Vagnozzi, E.~Giusarma, O.~Mena, K.~Freese, M.~Gerbino, S.~Ho and M.~Lattanzi,
  Unveiling $\nu$ secrets with cosmological data: neutrino masses and mass hierarchy,
  Phys.\ Rev.\ D {\bf 96}, no. 12, 123503 (2017).

\bibitem{Zhang:2017rbg}
  X.~Zhang,
  Weighing neutrinos in dynamical dark energy models,
  Sci.\ China Phys.\ Mech.\ Astron.\  {\bf 60}, no. 6, 060431 (2017).



\bibitem{Lorenz:2017fgo}
  C.~S.~Lorenz, E.~Calabrese and D.~Alonso,
  Distinguishing between neutrinos and time-varying dark energy through cosmic time,
  Phys.\ Rev.\ D {\bf 96}, no. 4, 043510 (2017).

\bibitem{Zhao:2017jma}
  M.~M.~Zhao, J.~F.~Zhang and X.~Zhang,
  Measuring growth index in a universe with massive neutrinos: a revisit of the general relativity test with the latest observations,
  Phys.\ Lett.\ B {\bf 779}, 473 (2018).

\bibitem{Vagnozzi:2018jhn}
  S.~Vagnozzi, S.~Dhawan, M.~Gerbino, K.~Freese, A.~Goobar and O.~Mena,
  Constraints on the sum of the neutrino masses in dynamical dark energy models with $w(z) \geq -1$ are tighter than those obtained in $\Lambda$CDM,
  arXiv:1801.08553 [astro-ph.CO].

\bibitem{Wang:2018lun}
  L.~F.~Wang, X.~N.~Zhang, J.~F.~Zhang and X.~Zhang,
  Impacts of gravitational-wave standard siren observation of the Einstein Telescope on weighing neutrinos in cosmology,
  Phys.\ Lett.\ B {\bf 782}, 87 (2018).


\bibitem{Li:2017iur}
  E.~K.~Li, H.~Zhang, M.~Du, Z.~H.~Zhou and L.~Xu,
  Probing the Neutrino Mass Hierarchy beyond $\Lambda$CDM Model,
  JCAP {\bf 1808}, 042 (2018)
  doi:10.1088/1475-7516/2018/08/042
  [arXiv:1703.01554 [astro-ph.CO]].

\bibitem{Wang:2017htc}
  S.~Wang, Y.~F.~Wang and D.~M.~Xia,
  Constraints on the sum of neutrino masses using cosmological data including the latest extended Baryon Oscillation Spectroscopic Survey DR14 quasar sample,
  Chin.\ Phys.\ C {\bf 42}, no. 6, 065103 (2018)
  doi:10.1088/1674-1137/42/6/065103
  [arXiv:1707.00588 [astro-ph.CO]].

\bibitem{Feng:2017usu}
  L.~Feng, J.~F.~Zhang and X.~Zhang,
  Search for sterile neutrinos in a universe of vacuum energy interacting with cold dark matter,
  Phys.\ Dark Univ.\ , 100261
  [Phys.\ Dark Univ.\  {\bf 23}, 100261 (2019)]
  doi:10.1016/j.dark.2018.100261
  [arXiv:1712.03148 [astro-ph.CO]].

\bibitem{Zhao:2016ecj}
  M.~M.~Zhao, Y.~H.~Li, J.~F.~Zhang and X.~Zhang,
  Constraining neutrino mass and extra relativistic degrees of freedom in dynamical dark energy models using Planck 2015 data in combination with low-redshift cosmological probes: basic extensions to ??CDM cosmology,
  Mon.\ Not.\ Roy.\ Astron.\ Soc.\  {\bf 469}, no. 2, 1713 (2017)
  doi:10.1093/mnras/stx978
  [arXiv:1608.01219 [astro-ph.CO]].

\bibitem{Zhang:2015uhk}
  X.~Zhang,
  Impacts of dark energy on weighing neutrinos after Planck 2015,
  Phys.\ Rev.\ D {\bf 93}, no. 8, 083011 (2016)
  doi:10.1103/PhysRevD.93.083011
  [arXiv:1511.02651 [astro-ph.CO]].

\bibitem{Huang:2015wrx}
  Q.~G.~Huang, K.~Wang and S.~Wang,
  Constraints on the neutrino mass and mass hierarchy from cosmological observations,
  Eur.\ Phys.\ J.\ C {\bf 76}, no. 9, 489 (2016)
  doi:10.1140/epjc/s10052-016-4334-z
  [arXiv:1512.05899 [astro-ph.CO]].

\bibitem{Wang:2016tsz}
  S.~Wang, Y.~F.~Wang, D.~M.~Xia and X.~Zhang,
  Impacts of dark energy on weighing neutrinos: mass hierarchies considered,
  Phys.\ Rev.\ D {\bf 94}, no. 8, 083519 (2016)
  doi:10.1103/PhysRevD.94.083519
  [arXiv:1608.00672 [astro-ph.CO]].



\bibitem{Vagnozzi:2019utt}
  S.~Vagnozzi,
  Cosmological searches for the neutrino mass scale and mass ordering,
  arXiv:1907.08010 [astro-ph.CO].

\bibitem{Vagnozzi:2019jzi}
  S.~Vagnozzi,
  Weigh them all! - Cosmological searches for the neutrino mass scale and mass ordering.

\bibitem{Giusarma:2016phn}
  E.~Giusarma, M.~Gerbino, O.~Mena, S.~Vagnozzi, S.~Ho and K.~Freese,
  Improvement of cosmological neutrino mass bounds,
  Phys.\ Rev.\ D {\bf 94}, no. 8, 083522 (2016)
  doi:10.1103/PhysRevD.94.083522
  [arXiv:1605.04320 [astro-ph.CO]].

\bibitem{Gariazzo:2018pei}
  S.~Gariazzo, M.~Archidiacono, P.~F.~de Salas, O.~Mena, C.~A.~Ternes and M.~Tš®rtola,
  Neutrino masses and their ordering: Global Data, Priors and Models,
  JCAP {\bf 1803}, 011 (2018)
  doi:10.1088/1475-7516/2018/03/011
  [arXiv:1801.04946 [hep-ph]].

\bibitem{Liu:2020vgn}
  Z.~Liu and H.~Miao,
  Neutrino mass and mass hierarchy in various dark energy,
  arXiv:2002.05563 [astro-ph.CO].

\bibitem{Choudhury:2018byy}
  S.~Roy Choudhury and S.~Choubey,
  Updated Bounds on Sum of Neutrino Masses in Various Cosmological Scenarios,
  JCAP {\bf 1809}, 017 (2018)
  doi:10.1088/1475-7516/2018/09/017
  [arXiv:1806.10832 [astro-ph.CO]].

\bibitem{Allahverdi:2016fvl}
  R.~Allahverdi, Y.~Gao, B.~Knockel and S.~Shalgar,
  Indirect Signals from Solar Dark Matter Annihilation to Long-lived Right-handed Neutrinos,
  Phys.\ Rev.\ D {\bf 95}, no. 7, 075001 (2017)
  doi:10.1103/PhysRevD.95.075001
  [arXiv:1612.03110 [hep-ph]].

\bibitem{Han:2017wnk}
  J.~Han, R.~Wang, W.~Wang and X.~N.~Wei,
  Neutrino mass matrices with one texture equality and one vanishing neutrino mass,
  Phys.\ Rev.\ D {\bf 96}, no. 7, 075043 (2017)
  doi:10.1103/PhysRevD.96.075043
  [arXiv:1705.05725 [hep-ph]].

\bibitem{Zhou:2014fva}
  X.~Y.~Zhou and J.~H.~He,
  Weighing neutrinos in $f(R)$ gravity in light of BICEP2,
  Commun.\ Theor.\ Phys.\  {\bf 62}, 102 (2014)
  doi:10.1088/0253-6102/62/1/18
  [arXiv:1406.6822 [astro-ph.CO]].

\bibitem{Huo:2011ve}
  Y.~Huo, T.~Li, Y.~Liao, D.~V.~Nanopoulos and Y.~Qi,
  Constraints on Neutrino Velocities Revisited,
  Phys.\ Rev.\ D {\bf 85}, 034022 (2012)
  doi:10.1103/PhysRevD.85.034022
  [arXiv:1112.0264 [hep-ph]].

\bibitem{Zhang:2019ipd}
  J.~F.~Zhang, B.~Wang and X.~Zhang,
  Forecast for weighing neutrinos in cosmology with SKA,
  Sci.\ China Phys.\ Mech.\ Astron.\  {\bf 63}, no. 8, 280411 (2020)
  doi:10.1007/s11433-019-1516-y
  [arXiv:1907.00179 [astro-ph.CO]].
  
\bibitem{DiazRivero:2019ukx}
  A.~Diaz Rivero, V.~Miranda and C.~Dvorkin,
  Observable Predictions for Massive-Neutrino Cosmologies with Model-Independent Dark Energy,
  Phys.\ Rev.\ D {\bf 100}, no. 6, 063504 (2019)
  doi:10.1103/PhysRevD.100.063504
  [arXiv:1903.03125 [astro-ph.CO]].  


\bibitem{Guo:2017hea}
  R.~Y.~Guo, Y.~H.~Li, J.~F.~Zhang and X.~Zhang,
  Weighing neutrinos in the scenario of vacuum energy interacting with cold dark matter: application of the parameterized post-Friedmann approach,
  JCAP {\bf 1705}, 040 (2017)
  doi:10.1088/1475-7516/2017/05/040
  [arXiv:1702.04189 [astro-ph.CO]].

\bibitem{Feng:2019jqa}
  L.~Feng, D.~Z.~He, H.~L.~Li, J.~F.~Zhang and X.~Zhang,
  Constraints on active and sterile neutrinos in an interacting dark energy cosmology,
  arXiv:1910.03872 [astro-ph.CO].

\bibitem{Guo:2018gyo}
  R.~Y.~Guo, J.~F.~Zhang and X.~Zhang,
  Exploring neutrino mass and mass hierarchy in the scenario of vacuum energy interacting with cold dark matte,
  Chin.\ Phys.\ C {\bf 42}, no. 9, 095103 (2018)
  doi:10.1088/1674-1137/42/9/095103
  [arXiv:1803.06910 [astro-ph.CO]].

\bibitem{Feng:2019mym}
  L.~Feng, H.~L.~Li, J.~F.~Zhang and X.~Zhang,
  Exploring neutrino mass and mass hierarchy in interacting dark energy models,
  Sci.\ China Phys.\ Mech.\ Astron.\  {\bf 63}, no. 2, 220401 (2020)
  doi:10.1007/s11433-019-9431-9
  [arXiv:1903.08848 [astro-ph.CO]].

\bibitem{Amendola:1999er}
  L.~Amendola,
  Coupled quintessence,
  Phys.\ Rev.\ D {\bf 62}, 043511 (2000)
  doi:10.1103/PhysRevD.62.043511
  [astro-ph/9908023].

\bibitem{Amendola:2001rc}
  L.~Amendola and D.~Tocchini-Valentini,
  Baryon bias and structure formation in an accelerating universe,
  Phys.\ Rev.\ D {\bf 66}, 043528 (2002)
  doi:10.1103/PhysRevD.66.043528
  [astro-ph/0111535].

\bibitem{Comelli:2003cv}
  D.~Comelli, M.~Pietroni and A.~Riotto,
  Dark energy and dark matter,
  Phys.\ Lett.\ B {\bf 571}, 115 (2003)
  doi:10.1016/j.physletb.2003.05.006
  [hep-ph/0302080].

\bibitem{Cai:2004dk}
  R.~G.~Cai and A.~Wang,
  Cosmology with interaction between phantom dark energy and dark matter and the coincidence problem,
  JCAP {\bf 0503}, 002 (2005)
  doi:10.1088/1475-7516/2005/03/002
  [hep-th/0411025].

\bibitem{Zhang:2005rg}
  X.~Zhang,
  Coupled quintessence in a power-law case and the cosmic coincidence problem,
  Mod.\ Phys.\ Lett.\ A {\bf 20}, 2575 (2005)
  doi:10.1142/S0217732305017597
  [astro-ph/0503072].

\bibitem{Zimdahl:2005bk}
  W.~Zimdahl,
  Interacting dark energy and cosmological equations of state,
  Int.\ J.\ Mod.\ Phys.\ D {\bf 14}, 2319 (2005)
  doi:10.1142/S0218271805007784
  [gr-qc/0505056].

\bibitem{Zhang:2004gc}
  X.~Zhang, F.~Q.~Wu and J.~Zhang,
  A New generalized Chaplygin gas as a scheme for unification of dark energy and dark matter,
  JCAP {\bf 0601}, 003 (2006)
  doi:10.1088/1475-7516/2006/01/003
  [astro-ph/0411221].

\bibitem{Wang:2006qw}
  B.~Wang, J.~Zang, C.~Y.~Lin, E.~Abdalla and S.~Micheletti,
  Interacting Dark Energy and Dark Matter: Observational Constraints from Cosmological Parameters,
  Nucl.\ Phys.\ B {\bf 778}, 69 (2007)
  doi:10.1016/j.nuclphysb.2007.04.037
  [astro-ph/0607126].

\bibitem{Guo:2007zk}
  Z.~K.~Guo, N.~Ohta and S.~Tsujikawa,
  Probing the Coupling between Dark Components of the Universe,
  Phys.\ Rev.\ D {\bf 76}, 023508 (2007)
  doi:10.1103/PhysRevD.76.023508
  [astro-ph/0702015 [ASTRO-PH]].

\bibitem{Bertolami:2007zm}
  O.~Bertolami, F.~Gil Pedro and M.~Le Delliou,
  Dark Energy-Dark Matter Interaction and the Violation of the Equivalence Principle from the Abell Cluster A586,
  Phys.\ Lett.\ B {\bf 654}, 165 (2007)
  doi:10.1016/j.physletb.2007.08.046
  [astro-ph/0703462 [ASTRO-PH]].

\bibitem{Zhang:2007uh}
  J.~Zhang, H.~Liu and X.~Zhang,
  Statefinder diagnosis for the interacting model of holographic dark energy,
  Phys.\ Lett.\ B {\bf 659}, 26 (2008)
  doi:10.1016/j.physletb.2007.10.086
  [arXiv:0705.4145 [astro-ph]].

\bibitem{Boehmer:2008av}
  C.~G.~Boehmer, G.~Caldera-Cabral, R.~Lazkoz and R.~Maartens,
  Dynamics of dark energy with a coupling to dark matter,
  Phys.\ Rev.\ D {\bf 78}, 023505 (2008)
  doi:10.1103/PhysRevD.78.023505
  [arXiv:0801.1565 [gr-qc]].

\bibitem{Valiviita:2008iv}
  J.~Valiviita, E.~Majerotto and R.~Maartens,
  Instability in interacting dark energy and dark matter fluids,
  JCAP {\bf 0807}, 020 (2008)
  doi:10.1088/1475-7516/2008/07/020
  [arXiv:0804.0232 [astro-ph]].

\bibitem{He:2008tn}
  J.~H.~He and B.~Wang,
  Effects of the interaction between dark energy and dark matter on cosmological parameters,
  JCAP {\bf 0806}, 010 (2008)
  doi:10.1088/1475-7516/2008/06/010
  [arXiv:0801.4233 [astro-ph]].

\bibitem{He:2009mz}
  J.~H.~He, B.~Wang and Y.~P.~Jing,
  Effects of dark sectors' mutual interaction on the growth of structures,
  JCAP {\bf 0907}, 030 (2009)
  doi:10.1088/1475-7516/2009/07/030
  [arXiv:0902.0660 [gr-qc]].

\bibitem{He:2009pd}
  J.~H.~He, B.~Wang and P.~Zhang,
  The Imprint of the interaction between dark sectors in large scale cosmic microwave background anisotropies,
  Phys.\ Rev.\ D {\bf 80}, 063530 (2009)
  doi:10.1103/PhysRevD.80.063530
  [arXiv:0906.0677 [gr-qc]].

\bibitem{Koyama:2009gd}
  K.~Koyama, R.~Maartens and Y.~S.~Song,
  Velocities as a probe of dark sector interactions,
  JCAP {\bf 0910}, 017 (2009)
  doi:10.1088/1475-7516/2009/10/017
  [arXiv:0907.2126 [astro-ph.CO]].

\bibitem{Xia:2009zzb}
  J.~Q.~Xia,
  Constraint on coupled dark energy models from observations,
  Phys.\ Rev.\ D {\bf 80}, 103514 (2009)
  doi:10.1103/PhysRevD.80.103514
  [arXiv:0911.4820 [astro-ph.CO]].

\bibitem{Li:2009zs}
  M.~Li, X.~D.~Li, S.~Wang, Y.~Wang and X.~Zhang,
  Probing interaction and spatial curvature in the holographic dark energy model,
  JCAP {\bf 0912}, 014 (2009)
  doi:10.1088/1475-7516/2009/12/014
  [arXiv:0910.3855 [astro-ph.CO]].

\bibitem{Zhang:2009qa}
  L.~Zhang, J.~Cui, J.~Zhang and X.~Zhang,
  Interacting model of new agegraphic dark energy: Cosmological evolution and statefinder diagnostic,
  Int.\ J.\ Mod.\ Phys.\ D {\bf 19}, 21 (2010)
  doi:10.1142/S0218271810016245
  [arXiv:0911.2838 [astro-ph.CO]].

\bibitem{Wei:2010cs}
  H.~Wei,
  Cosmological Constraints on the Sign-Changeable Interactions,
  Commun.\ Theor.\ Phys.\  {\bf 56}, 972 (2011)
  doi:10.1088/0253-6102/56/5/29
  [arXiv:1010.1074 [gr-qc]].

\bibitem{Li:2010ak}
  Y.~Li, J.~Ma, J.~Cui, Z.~Wang and X.~Zhang,
  Interacting model of new agegraphic dark energy: observational constraints and age problem,
  Sci.\ China Phys.\ Mech.\ Astron.\  {\bf 54}, 1367 (2011)
  doi:10.1007/s11433-011-4382-1
  [arXiv:1011.6122 [astro-ph.CO]].

\bibitem{He:2010im}
  J.~H.~He, B.~Wang and E.~Abdalla,
  Testing the interaction between dark energy and dark matter via latest observations,
  Phys.\ Rev.\ D {\bf 83}, 063515 (2011)
  doi:10.1103/PhysRevD.83.063515
  [arXiv:1012.3904 [astro-ph.CO]].

\bibitem{Li:2011ga}
  Y.~H.~Li and X.~Zhang,
  Running coupling: Does the coupling between dark energy and dark matter change sign during the cosmological evolution?,
  Eur.\ Phys.\ J.\ C {\bf 71}, 1700 (2011)
  doi:10.1140/epjc/s10052-011-1700-8
  [arXiv:1103.3185 [astro-ph.CO]].

\bibitem{Fu:2011ab}
  T.~F.~Fu, J.~F.~Zhang, J.~Q.~Chen and X.~Zhang,
  Holographic Ricci dark energy: Interacting model and cosmological constraints,
  Eur.\ Phys.\ J.\ C {\bf 72}, 1932 (2012)
  doi:10.1140/epjc/s10052-012-1932-2
  [arXiv:1112.2350 [astro-ph.CO]].

\bibitem{Zhang:2012uu}
  Z.~Zhang, S.~Li, X.~D.~Li, X.~Zhang and M.~Li,
  Revisit of the Interaction between Holographic Dark Energy and Dark Matter,
  JCAP {\bf 1206}, 009 (2012)
  doi:10.1088/1475-7516/2012/06/009
  [arXiv:1204.6135 [astro-ph.CO]].

\bibitem{Zhang:2013lea}
  J.~Zhang, L.~Zhao and X.~Zhang,
  Revisiting the interacting model of new agegraphic dark energy,
  Sci.\ China Phys.\ Mech.\ Astron.\  {\bf 57}, 387 (2014)
  doi:10.1007/s11433-013-5378-9
  [arXiv:1306.1289 [astro-ph.CO]].

\bibitem{Li:2013bya}
  Y.~H.~Li and X.~Zhang,
  Large-scale stable interacting dark energy model: Cosmological perturbations and observational constraints,
  Phys.\ Rev.\ D {\bf 89}, no. 8, 083009 (2014)
  doi:10.1103/PhysRevD.89.083009
  [arXiv:1312.6328 [astro-ph.CO]].

\bibitem{Geng:2015ara}
  J.~J.~Geng, Y.~H.~Li, J.~F.~Zhang and X.~Zhang,
  Redshift drift exploration for interacting dark energy,
  Eur.\ Phys.\ J.\ C {\bf 75}, no. 8, 356 (2015)
  doi:10.1140/epjc/s10052-015-3581-8
  [arXiv:1501.03874 [astro-ph.CO]].

\bibitem{Yin:2015pqa}
  J.~L.~Cui, L.~Yin, L.~F.~Wang, Y.~H.~Li and X.~Zhang,
  A closer look at interacting dark energy with statefinder hierarchy and growth rate of structure,
  JCAP {\bf 1509}, 024 (2015)
  doi:10.1088/1475-7516/2015/09/024
  [arXiv:1503.08948 [astro-ph.CO]].

\bibitem{Murgia:2016ccp}
  R.~Murgia, S.~Gariazzo and N.~Fornengo,
  Constraints on the Coupling between Dark Energy and Dark Matter from CMB data,
  JCAP {\bf 1604}, 014 (2016)
  doi:10.1088/1475-7516/2016/04/014
  [arXiv:1602.01765 [astro-ph.CO]].

\bibitem{Wang:2016lxa}
  B.~Wang, E.~Abdalla, F.~Atrio-Barandela and D.~Pavon,
  Dark Matter and Dark Energy Interactions: Theoretical Challenges, Cosmological Implications and Observational Signatures,
  Rept.\ Prog.\ Phys.\  {\bf 79}, no. 9, 096901 (2016)
  doi:10.1088/0034-4885/79/9/096901
  [arXiv:1603.08299 [astro-ph.CO]].

\bibitem{Pourtsidou:2016ico}
  A.~Pourtsidou and T.~Tram,
  Reconciling CMB and structure growth measurements with dark energy interactions,
  Phys.\ Rev.\ D {\bf 94}, no. 4, 043518 (2016)
  doi:10.1103/PhysRevD.94.043518
  [arXiv:1604.04222 [astro-ph.CO]].

\bibitem{Costa:2016tpb}
  A.~A.~Costa, X.~D.~Xu, B.~Wang and E.~Abdalla,
  Constraints on interacting dark energy models from Planck 2015 and redshift-space distortion data,
  JCAP {\bf 1701}, 028 (2017)
  doi:10.1088/1475-7516/2017/01/028
  [arXiv:1605.04138 [astro-ph.CO]].

\bibitem{Sola:2016ecz}
  J.~Sol? Peracaula, J.~de Cruz P??rez and A.~G??mez-Valent,
  Dynamical dark energy vs. $\Lambda$ = const in light of observations,
  EPL {\bf 121}, no. 3, 39001 (2018)
  doi:10.1209/0295-5075/121/39001
  [arXiv:1606.00450 [gr-qc]].

\bibitem{Feng:2016djj}
  L.~Feng and X.~Zhang,
  Revisit of the interacting holographic dark energy model after Planck 2015,
  JCAP {\bf 1608}, 072 (2016)
  doi:10.1088/1475-7516/2016/08/072
  [arXiv:1607.05567 [astro-ph.CO]].

\bibitem{Xia:2016vnp}
  D.~M.~Xia and S.~Wang,
  Constraining interacting dark energy models with latest cosmological observations,
  Mon.\ Not.\ Roy.\ Astron.\ Soc.\  {\bf 463}, no. 1, 952 (2016)
  doi:10.1093/mnras/stw2073
  [arXiv:1608.04545 [astro-ph.CO]].

\bibitem{vandeBruck:2016hpz}
  C.~van de Bruck, J.~Mifsud and J.~Morrice,
  Testing coupled dark energy models with their cosmological background evolution,
  Phys.\ Rev.\ D {\bf 95}, no. 4, 043513 (2017)
  doi:10.1103/PhysRevD.95.043513
  [arXiv:1609.09855 [astro-ph.CO]].

\bibitem{Sola:2016zeg}
  J.~Sol?,
  Cosmological constant vis-a-vis dynamical vacuum: bold challenging the $\Lambda$CDM,
  Int.\ J.\ Mod.\ Phys.\ A {\bf 31}, no. 23, 1630035 (2016)
  doi:10.1142/S0217751X16300350
  [arXiv:1612.02449 [astro-ph.CO]].

\bibitem{Kumar:2017dnp}
  S.~Kumar and R.~C.~Nunes,
  Echo of interactions in the dark sector,
  Phys.\ Rev.\ D {\bf 96}, no. 10, 103511 (2017)
  doi:10.1103/PhysRevD.96.103511
  [arXiv:1702.02143 [astro-ph.CO]].

\bibitem{Sola:2017jbl}
  J.~Sol? Peracaula, J.~d.~C.~Perez and A.~Gomez-Valent,
  Possible signals of vacuum dynamics in the Universe,
  Mon.\ Not.\ Roy.\ Astron.\ Soc.\  {\bf 478}, no. 4, 4357 (2018)
  doi:10.1093/mnras/sty1253
  [arXiv:1703.08218 [astro-ph.CO]].



\bibitem{Li:2014eha}
  Y.~H.~Li, J.~F.~Zhang and X.~Zhang,
  Parametrized Post-Friedmann Framework for Interacting Dark Energy,
  Phys.\ Rev.\ D {\bf 90}, no. 6, 063005 (2014)
  doi:10.1103/PhysRevD.90.063005
  [arXiv:1404.5220 [astro-ph.CO]].

\bibitem{Li:2014cee}
  Y.~H.~Li, J.~F.~Zhang and X.~Zhang,
  Exploring the full parameter space for an interacting dark energy model with recent observations including redshift-space distortions: Application of the parametrized post-Friedmann approach,
  Phys.\ Rev.\ D {\bf 90}, no. 12, 123007 (2014)
  doi:10.1103/PhysRevD.90.123007
  [arXiv:1409.7205 [astro-ph.CO]].

\bibitem{Li:2015vla}
  Y.~H.~Li, J.~F.~Zhang and X.~Zhang,
  Testing models of vacuum energy interacting with cold dark matter,
  Phys.\ Rev.\ D {\bf 93}, no. 2, 023002 (2016)
  doi:10.1103/PhysRevD.93.023002
  [arXiv:1506.06349 [astro-ph.CO]].

\bibitem{Zhang:2017ize}
  X.~Zhang,
  Probing the interaction between dark energy and dark matter with the parametrized post-Friedmann approach,
  Sci.\ China Phys.\ Mech.\ Astron.\  {\bf 60}, no. 5, 050431 (2017)
  doi:10.1007/s11433-017-9013-7
  [arXiv:1702.04564 [astro-ph.CO]].

\bibitem{Feng:2018yew}
  L.~Feng, Y.~H.~Li, F.~Yu, J.~F.~Zhang and X.~Zhang,
  Exploring interacting holographic dark energy in a perturbed universe with parameterized post-Friedmann approach,
  Eur.\ Phys.\ J.\ C {\bf 78}, no. 10, 865 (2018)
  doi:10.1140/epjc/s10052-018-6338-3
  [arXiv:1807.03022 [astro-ph.CO]].

\bibitem{Hu:2008zd}
  W.~Hu,
  Parametrized Post-Friedmann Signatures of Acceleration in the CMB,
  Phys.\ Rev.\ D {\bf 77}, 103524 (2008)
  doi:10.1103/PhysRevD.77.103524
  [arXiv:0801.2433 [astro-ph]].

\bibitem{Fang:2008sn}
  W.~Fang, W.~Hu and A.~Lewis,
  Crossing the Phantom Divide with Parameterized Post-Friedmann Dark Energy,
  Phys.\ Rev.\ D {\bf 78}, 087303 (2008)
  doi:10.1103/PhysRevD.78.087303
  [arXiv:0808.3125 [astro-ph]].

\bibitem{Lewis:2002ah}
  A.~Lewis and S.~Bridle,
  Cosmological parameters from CMB and other data: A Monte Carlo approach,
  Phys.\ Rev.\ D {\bf 66}, 103511 (2002)
  doi:10.1103/PhysRevD.66.103511
  [astro-ph/0205436].

\bibitem{Gelman:1992zz}
  A.~Gelman and D.~B.~Rubin,
  Inference from Iterative Simulation Using Multiple Sequences,
  Statist.\ Sci.\  {\bf 7}, 457 (1992).
  doi:10.1214/ss/1177011136

\bibitem{Aghanim:2018eyx}
  N.~Aghanim {\it et al.} [Planck Collaboration],
  Planck 2018 results. VI. Cosmological parameters,
  arXiv:1807.06209 [astro-ph.CO].


\bibitem{Beutler:2011hx}
  F.~Beutler {\it et al.},
  The 6dF Galaxy Survey: Baryon Acoustic Oscillations and the Local Hubble Constant,
  Mon.\ Not.\ Roy.\ Astron.\ Soc.\  {\bf 416}, 3017 (2011)
  doi:10.1111/j.1365-2966.2011.19250.x
  [arXiv:1106.3366 [astro-ph.CO]].

\bibitem{Ross:2014qpa}
  A.~J.~Ross, L.~Samushia, C.~Howlett, W.~J.~Percival, A.~Burden and M.~Manera,
  The clustering of the SDSS DR7 main Galaxy sample C I. A 4 per cent distance measure at $z = 0.15$,
  Mon.\ Not.\ Roy.\ Astron.\ Soc.\  {\bf 449}, no. 1, 835 (2015)
  doi:10.1093/mnras/stv154
  [arXiv:1409.3242 [astro-ph.CO]].

\bibitem{Alam:2016hwk}
  S.~Alam {\it et al.} [BOSS Collaboration],
  The clustering of galaxies in the completed SDSS-III Baryon Oscillation Spectroscopic Survey: cosmological analysis of the DR12 galaxy sample,
  Mon.\ Not.\ Roy.\ Astron.\ Soc.\  {\bf 470}, no. 3, 2617 (2017)
  doi:10.1093/mnras/stx721
  [arXiv:1607.03155 [astro-ph.CO]].

\bibitem{Scolnic:2017caz}
  D.~M.~Scolnic {\it et al.},
  The Complete Light-curve Sample of Spectroscopically Confirmed SNe Ia from Pan-STARRS1 and Cosmological Constraints from the Combined Pantheon Sample,
  Astrophys.\ J.\  {\bf 859}, no. 2, 101 (2018)
  doi:10.3847/1538-4357/aab9bb
  [arXiv:1710.00845 [astro-ph.CO]].

\bibitem{Riess:2019cxk}
  A.~G.~Riess, S.~Casertano, W.~Yuan, L.~M.~Macri and D.~Scolnic,
  Large Magellanic Cloud Cepheid Standards Provide a 1\% Foundation for the Determination of the Hubble Constant and Stronger Evidence for Physics beyond $\Lambda$CDM,
  Astrophys.\ J.\  {\bf 876}, no. 1, 85 (2019)
  doi:10.3847/1538-4357/ab1422
  [arXiv:1903.07603 [astro-ph.CO]].


\end{thebibliography}
\end{document}